\documentclass[11pt]{article}
\usepackage{jheppub}
\usepackage{graphicx}
\usepackage{amsmath, amssymb}
\usepackage{tikz}
\usepackage{bm}

\graphicspath{{./Figures/}}

\newcommand{\be}{\begin{eqnarray}}
\newcommand{\ee}{\end{eqnarray}}

\newcommand{\bn}{\begin{enumerate}}
\newcommand{\en}{\end{enumerate}}

\newcommand{\bDelta}{\overline{\Delta}}
\newcommand{\bbDelta}{\overline{\bm{\Delta}}}

\usetikzlibrary{arrows}
\tikzstyle{every picture}+=[remember picture]
\tikzstyle{na} = [baseline=-.5ex]

\def\CA{{\cal A}}

\def\CI{{\cal I}}

\def\bfA{{\bf A}}
\def\bfB{{\bf B}}

\def\tq{\tilde{q}}
\def\ty{\tilde{y}}

\def\half{\frac{1}{2}}

\def\tr{{\rm tr}}

\usetikzlibrary{arrows}
\usetikzlibrary{positioning}
\usetikzlibrary{decorations.pathmorphing}
\usetikzlibrary{decorations.markings}
\def\arrowhead{angle 90}
\tikzset{>=\arrowhead}
\pgfarrowsdeclaredouble{<<}{>>}{\arrowhead}{\arrowhead}
\pgfarrowsdeclaretriple{<<<}{>>>}{\arrowhead}{\arrowhead}
\pgfarrowsdeclaredouble{<<<<}{>>>>}{<<}{>>}
\tikzstyle{G}=[draw, circle, minimum size=1em, scale=1, inner sep=2pt]
\tikzstyle{B}=[draw,circle,fill=black,scale=1]
\tikzstyle{H}=[draw,circle,fill=white,scale=1]
\tikzstyle{F}=[draw, rectangle, minimum width=1em, minimum height=1em, scale=1]
\tikzstyle{every picture}=[scale=1,baseline=(current bounding box.south)]

\newcommand{\bS}{\ensuremath{\mathbb{S}}}
\newcommand{\bW}{\ensuremath{\mathbb{W}}}

\newcommand{\scI}{\ensuremath{\mathcal{I}}}

\newcommand{\matWs}[5]{\ensuremath{
\mathcal{R}(#1)
\left[\begin{array}{cc}
#5 & #4 \\
#2 & #3 \\
\end{array}
\right]
}}

\newcommand{\matWd}[8]{\ensuremath{
\mathcal{R}
\left(\begin{array}{cc}
#4 & #3 \\
#1 & #2 \\
\end{array}
\right)
\left[\begin{array}{cc}
#8 & #7 \\
#5 & #6 \\
\end{array}
\right]
}}

\newcommand{\beq}{\begin{equation}\begin{aligned}}
\newcommand{\eeq}{\end{aligned}\end{equation}}

\title{Integrability from 2d $\mathcal{N}=(2,2)$ Dualities}

\author[a,b]{Masahito Yamazaki}
\author[c]{and Wenbin Yan}

\affiliation[a]{Kavli Institute for the Physics and Mathematics of the Universe (WPI),\\
University of Tokyo, Chiba 277-8583, Japan}
\affiliation[b]{School of Natural Sciences, Institute for Advanced Study, Princeton, NJ 08540, USA}
\affiliation[c]{Walter Burke Institute for Theoretical Physics,\\
California Institute of Technology, 452-48, Pasadena, CA 91125
}

\emailAdd{masahito.yamazaki@ipmu.jp}
\emailAdd{wbyan@theory.caltech.edu}

\preprint{IPMU15-0051, CALT-TH-2015-022}

\abstract{
We study integrable models in the context of the recently discovered Gauge/YBE correspondence,
where the Yang-Baxter equation is promoted to a duality between two supersymmetric gauge theories. We study 
flavored elliptic genus of 2d $\mathcal{N}=(2,2)$ quiver gauge theories, which theories are defined from statistical lattices regarded as quiver diagrams. Our R-matrices are written in terms of theta functions, and
simplifies considerably when the gauge groups at the quiver nodes are Abelian.
We also discuss the modularity properties of the R-matrix,
reduction of 2d index to 1d Witten index, and string theory realizations of our theories.
}

\begin{document}
\maketitle
\flushbottom

\section{Introduction}

In this paper we discuss the celebrated Yang-Baxter equation (YBE),
which is the one of the most fundamental characterizations of integrable models (see e.g.\ \cite{Baxter:1982zz} and references therein).
YBE has several different expressions. For concreteness let us here use the following version,
formulated in the language of Interaction-Round-a-Face (IRF) model
(we will comment on the formulation as a vertex model later):
\begin{equation}
\begin{split}
& \sum_g \matWs{u}{a}{b}{g}{f}
\matWs{u+v}{g}{b}{c}{d}
\matWs{v}{f}{g}{d}{e}
\\
& \qquad \qquad=\sum_g
 \matWs{v}{a}{b}{c}{g}
 \matWs{u+v}{f}{a}{g}{e}
\matWs{u}{g}{c}{d}{e} \ .
\end{split}
\label{eq.YBE}
\end{equation}
Here $\mathcal{R}(u)$ is known as the {\it R-matrix},
where the parameters $u, v$ are continuous parameters called {\it spectral parameters}.
Each 
\begin{align}
\mathcal{R}(u)=\matWs{u}{a}{b}{c}{d}
\label{R_IRF}
\end{align}
has four indices: $a, b, c, d$.
The indices $a, b, \ldots$ run over the possible values the spins could
take in a given integrable model;
for example in the Ising model we have $a, b, \ldots =\pm 1 \in \mathbb{Z}_2:=\mathbb{Z}/2\mathbb{Z}$.

\begin{figure}[htbp]
\centering\includegraphics[scale=0.5]{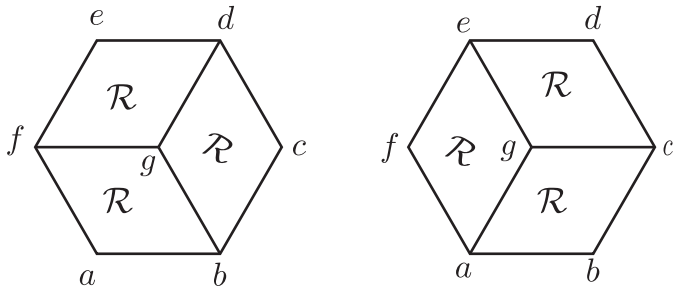}
\label{fig.YBE}
\caption{Graphical representation of YBE. A paralellogram represents the R-matrix, which for an IRF model is the weight $\mathcal{R}(F)$ for a face $F$. The orientation of the letter $\mathcal{R}$ on a face represents the index structure of the associated R-matrix.
Note that the R-matrix \eqref{R_IRF}
 in general is not symmetric under the cyclic permutation of its indices $a,b,c,d$.}
\end{figure}

A graphical representation of YBE is given in Figure \ref{fig.YBE}.
Here an R-matrix is represented by a parallelogram, whose four vertices are associated with the four indices
$a, b, c, d$ of the R-matrix $\matWs{u}{a}{b}{c}{d}$.

Given a solution to YBE \eqref{eq.YBE}, we can define
an integrable model by associating a
Boltzmann weight $\mathcal{R}(F)$ to a face $F$, and
by summing over all the possible spin configurations:
\beq
Z_{\textrm{2d integrable spin}}=\sum_{\rm spins} \prod_{F\textrm{: face}} \mathcal{R}(F) \ .
\label{ZF}
\eeq
The YBE ensures that the transfer matrices commute (under appropriate boundary conditions\footnote{For example, we can choose fixed or periodic boundary conditions.}),
and when transfer matrices are expanded with respect to spectral parameters we obtain
an infinite set of conserved charges.

The Yang-Baxter equation is a highly over-constrained equation.
For example, suppose that the spins take values in $\mathbb{Z}_N:=\mathbb{Z}/N \mathbb{Z}$, with $N$ large.
Then the R-matrix has $\mathcal{O}(N^2)$ variables, whereas the constraints from the Yang-Baxter equation grows as $\mathcal{O}(N^3)$. This problem becomes more severe for integrable models with continuous spins (our model below fits to this category),
which can be formally thought of as the limit $N\to \infty$, and it looks almost impossible to find any solution at all.
Despite these naive expectations, people have found a variety of solutions of YBE.
The natural question is then {\it why integrable models exist}.

\bigskip

Recently there is a new look on this long-standing problem \cite{Yamazaki:2013nra} (see also  \cite{Yamazaki:2012cp,Terashima:2012cx}; this correspondence is called the {\it Gauge/YBE correspondence}.). Namely Yang-Baxter equation is
promoted (``categorified'') to a duality (Yang-Baxter duality) between two supersymmetric quiver gauge theories,
which duality in turn follows from a sequence of 4d $\mathcal{N}=1$ Seiberg dualities. The basic logic is that this duality automatically generates a mathematical equality of the
partition functions of the two theories, and the YBE follows directly from the Yang-Baxter duality.

The lift of the YBE to a duality is rather powerful, since duality is not about a single identity
but rather a set of such identities---namely we can compute various partition functions and observables (satisfying certain constraints to be discussed later), and each of these gives rise to (in general) different solutions
for the YBE. For example, in \cite{Yamazaki:2013nra} the 4d lens index \cite{Benini:2011nc}, the twisted partition function on $S^1\times S^3/\mathbb{Z}_r$, gave rise to a large class of integrable models which are previously unknown in the literature.  We can also discuss various degenerations of the model.
For $r=1$ the model reduces to the ``master solution'' of \cite{Bazhanov:2011mz} (see also \cite{Bazhanov:2010kz,Spiridonov:2010em,Bazhanov:2013bh}), where the name ``master'' originates from the fact that it successfully reproduced all the known solutions of the star-triangle relations with 
positive Boltzmann weights. The degeneration of this model gives rise to a variety of integrable models \cite{Bazhanov:2007mh,Bazhanov:2010kz,Bazhanov:2011mz,Volkov:1992uv,FaddeevVolkovAbelian}, eventually all the way down to the Ising model, which is at the bottom of this hierarchy of integrable models.

\bigskip

The natural question is whether we can adopt the same logic to supersymmetric field theories in
dimensions other than four. In \cite{Yamazaki:2013nra} it has already been pointed out that
a similar story works for (for example, supersymmetric $S^1\times S^2$ partition function of) 3d $\mathcal{N}=2$ version of Seiberg duality (Aharony duality \cite{Aharony:1997gp}); 
as far as the underlying combinatorial structure of the quiver is the same, we should obtain a solution to the Yang-Baxter equation.

The goal of this paper is to generalize the logic of \cite{Yamazaki:2013nra} to 2d $\mathcal{N}=(2,2)$ quiver gauge theories\footnote{Towards the completion of this paper we received \cite{Yagi:2015lha}, which also discusses integrable models associated with 2d $\mathcal{N}=(2,2)$ theories.}.
With the help of the Seiberg-like duality in 2d, we propose a 2d version of the
Yang-Baxter duality, and compute supersymmetric partition function on
$T^2$ (elliptic genus) \cite{Gadde:2013dda,Benini:2013nda,Benini:2013xpa}.
This automatically gives rise to 2d classical integrable models:
\beq
\scI_\textrm{2d $\mathcal{N}=(2,2)$ theory}[
T^2]= Z_\textrm{2d integrable spin}\ .
\label{eq.GaugeYBE}
\eeq
The resulting 2d spin system obeys either the periodic or fixed boundary condition,
depending on whether we have a quiver diagram on a torus or on a plane.
Correspondingly the quiver diagram is drawn either on a torus \cite{Franco:2005rj} or on a disc \cite{Franco:2012mm,Xie:2012mr}.

\bigskip

While this is to some extent a simple adoption of existing 4d techniques to 2d,
we encounter some new features. Along the way we will
also clarify some aspects of the Gauge/YBE correspondence itself.

The first subtlety is that the $T^2$ partition function has a subtlety in the choice of the contour of integration,
which should be taken into account in the discussion of integrable models.
Second, the R-matrices are written in terms of the well-known theta functions,
and we can {\it directly prove} the YBE by evaluating the integrals. This would be helpful to
those mathematical physicists
who do not
wish to go through the derivation from non-perturbative dualities in gauge theories.
This contrasts with the case of the 4d, where the invariance of the $S^1\times S^3/\mathbb{Z}_r$ index for a Seiberg-dual pair is
not proven mathematically.\footnote{For the case of the $S^1\times S^3$ partition function ($r=1$ in our previous notation), this follows from the identity of \cite{RainsTransf}, as pointed out in \cite{Dolan:2008qi}.}
Third, for the case of the $T^2$ partition function, our R-matrix (and hence the resulting integrable model) 
depends on a modular parameter $\tau$ of the torus, and has a nice modular property
under the $SL(2, \mathbb{Z})$ action on $\tau$.\footnote{See, however, discussion of the $SL(3, \mathbb{Z})$ modularity of the 4d $S^1\times S^3$ index \cite{Spiridonov:2012ww}.}

\bigskip
The rest of this paper is organized as follows. In section \ref{sec.2} we review the basic logic of the Gauge/YBE correspondence,
and apply it to the case of 2d $\mathcal{N}=(2,2)$ Seiberg-like duality.
Along the way we highlight some of the key ingredients which are crucial for the discussion of this paper.
In section
\ref{sec:indexdef} 
we explicitly construct the integrable model
corresponding to $T^2$. We also include comments on brane realizations in section 
\ref{sec.brane_realization}.
The final section (section \ref{sec.conclusion}) contains concluding remarks,
with technical material summarized in the two appendices.

\section{Yang-Baxter Equation from Duality} \label{sec.2}

In this section we summarize the basic logic of how to associate an integrable model
to a 2d Seiberg-like duality.
The explanation of the Gauge/YBE correspondence here closely follows \cite{Yamazaki:2013nra},
however the presentation here is improved in a number of technical points. 
We also emphasize several key aspects crucial for the 2d duality,
which are implicit in the 4d discussion of \cite{Yamazaki:2013nra}.

\subsection{Seiberg-like Duality}

In the construction of \cite{Yamazaki:2013nra},
the crucial input for the 4d Yang-Baxter duality was the 4d Seiberg duality \cite{Seiberg:1994pq}.
We will therefore look for a Seiberg-like duality for 2d
$\mathcal{N}=(2,2)$ theories.
There are several different versions of Seiberg-like dualities for 2d $\mathcal{N}=(2,2)$ theories in the literature, starting with the seminal work of \cite{Hori:2006dk}.
For our purposes we need a version \cite{Benini:2012ui,Gadde:2013dda}
whose matter contents are essentially the same as that of the 4d $\mathcal{N}=1$ Seiberg duality. Let us quickly summarize this duality.

The 2d $\mathcal{N}=(2,2)$ theory is a gauged linear sigma model (GLSM),
and its Lagrangian can be obtained from the
dimensional reduction of the parent 4d $\mathcal{N}=1$ theory.
The definition of the chiral and vector multiplets, for example, works in a similar manner.

There are several differences, however. One difference is that we can define
a twisted superfield
$\Sigma:=\overline{D}_{+} D_{-} V$, satisfying
$\overline{D}_{+} \Sigma=D_{-} \Sigma=0$.
We can furthermore consider the twisted superpotential
\begin{align}
\mathcal{L}_{\rm FI}=
\frac{1}{2}
\left(
  -t \int \!d^2  \tilde{\theta}\, \, \Sigma + \textrm{(c.c.)}
\right)
= -r D + \theta F_{01} \ ,
\label{eq.complexFI}
\end{align}
where we defined the complexified FI parameter $t$ by
\begin{align}
t:=r+ i\theta  \ .
\end{align}
Here $\theta$ angle is periodic, $\theta\sim \theta+2\pi$, and it hence it is natural to
consider the exponentiated single-valued variable
\begin{align}
z:=(-1)^N e^{t} \ .
\end{align}
Here we take the gauge group to be $U(N_c\equiv N)$.
Note that we include the diagonal $U(1)$ factor in the gauge group,
which plays a crucial role in the dynamics of 2d $\mathcal{N}=(2,2)$ theories;
this sharply contrasts with the case of 4d $\mathcal{N}=1$ theory, where the $U(1)$ factor decouples.

On one side of the duality (electric theory) we have $U(N)$ gauge theory with $N_f$ flavors, i.e.\ $N_f$ fundamentals $q_i$ and $N_f$ antifundamentals $\tilde{q}^i$.
Here $i=1, \ldots, N_f$ is the index for the flavor symmetry.
We have no superpotential:
$W_{\rm electric}=0$.
We also turn on the twisted superpotential \eqref{eq.complexFI} with
complexified FI parameter $z_{\rm electric}$.

On the other side (magnetic theory)
we have $U(N_f-N)$ gauge theory with $N_f$ flavors,
i.e.\ $N_f$ fundamentals $Q_i$ and $N_f$ antifundamentals $\tilde{Q}^i$.
We in addition have a meson field $M_i{}^j$ with superpotential coupling
\begin{align}
W_{\rm magnetic}=\textrm{Tr}(\tilde{Q}^i M_i{}^j Q_j) \ .
\label{W_mag}
\end{align}
We also turn on the twisted superpotential \eqref{eq.complexFI} with
complexified FI parameter $z_{\rm magnetic}$.

The statement for the duality is that the electric and magnetic theories flow in the IR
to the same fixed point, if we identify
\begin{align}
z_{\rm electric}= z_{\rm magnetic}^{-1} \ .
\label{z_inverse}
\end{align}
It turns out that this is part of the transformation properties of the cluster $y$-variable \cite{Benini:2014mia}.

In the literature there are several consistency checks for this duality \cite{Gadde:2013dda,Benini:2013nda,Benini:2013xpa}.
It is shown that they have the same chiral ring as well as the twisted chiral ring.
They also have the same $S^2$ partition function and the $T^2$ partition function,
and finally the duality has a geometrical counterpart in the geometry of the Grassmannian $\textrm{Gr}(N, N_f)$.

This duality  can be checked by explicitly computing the
$T^2$ partition functions,
which in turn ensures the YBE.
Note that our construction of integrable models in itself does not rely on the
full non-perturbative duality, but rather on the equality of the specific 
partition function (namely $T^2$ partition function).
However the underlying gauge theory duality is the ultimate reason
why these integrable models should exist.

\bigskip

\begin{figure}[htbp]
\centering\includegraphics[scale=0.5]{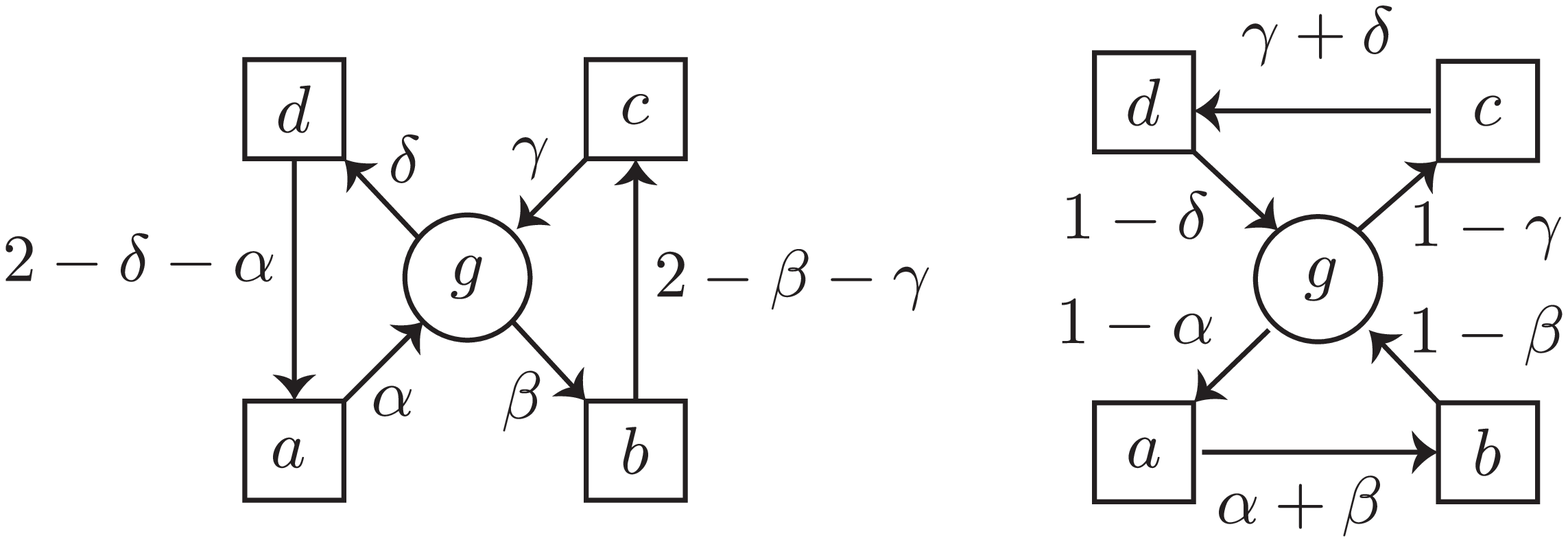}
  \caption{The change of the quiver diagram for the 2d Seiberg-like duality.
  A circle (box) represents a gauge (global) symmetry.
  In the following we specialize to the case $N_a=N_b=N_c=N_d=N_g=N$ (and hence $N_f=2N$ for the gauge group in the middle). The parameters $\alpha, \beta, \cdots$ represents the R-charge of the bifundamental fields. For a closed loop (triangle) their R-charges sum up to two.
  }
  \label{fig.2dSeiberg_quiver}
\end{figure}

For our purposes of constructing integrable models, what is important here is that the matter content for the 2d Seiberg-like duality
 is essentially the same as the
4d Seiberg duality
This means that we can immediately borrow most of the results from
\cite{Yamazaki:2013nra}.
For example, this Seiberg-like duality for a quiver gauge theory is represented graphically as in
Figure \ref{fig.2dSeiberg_quiver}.
As explained in \cite{Yamazaki:2013nra}, this is the gauge theory counterpart of the relation known as the
star-star relation \cite{Baxter:1986,Bazhanov:1992jqa}  in integrable models.
The star-star relation ensures the YBE\footnote{However, only a limit subset of the solutions to the YBE originates from the star-star relation.}; correspondingly
2d Seiberg duality implies the 2d version of the Yang-Baxter duality.

For simplicity of the presentation in the following 
we will concentrate on the case $N_f=2N$, with 
all the quiver nodes have gauge group $U(N)$ with the same rank $N$.
One advantage of this choice is that the rank of the gauge group,
and hence number of the components of the spin of the integrable model at a lattice site, 
is preserved by the Seiberg-like duality. 

\subsection{Constraints on Partition Functions}

Let us here be more precise as to which partition functions
we could consider in \eqref{eq.GaugeYBE}.
Closer inspection reveals that the logic of \cite{Yamazaki:2013nra}
relies crucially on the following four general properties of the partition function.

\paragraph{Invariance under the IR duality.}

The first requirement is that the partition function coincides for two UV dual theories both of which flow to the same IR fixed point.
This ensures the invariance of the partition function under the Yang-Baxter duality.
Note that this is the case if the partition function is independent of the gauge coupling constant,
which is indeed the case for $T^2$
partition function
discussed in this paper\footnote{Even if the partition function
depends non-trivially on the gauge coupling constant, we should be able to extract a precise mathematical identity from a gauge theory duality, as long as we can keep track of the dependence of the partition function on the gauge coupling constant.}.

The invariance sometimes holds up to overall constant factors, or up to the appropriate change of parameters. In these cases a care is needed for the identification of a precise mathematical identity.

\paragraph{Factorization}

Let us consider the partition function for a quiver gauge theory,
with gauge field (matter) associated with the vertex (edge) of the quiver diagram.
Then the second requirement is that the
classical as well as 1-loop contribution to the integrand of the integral/sum
expression for the partition function
partition function for this quiver gauge theory
factories into contributions from gauge field and matters.
Schematically,
\begin{align}
Z_{\rm quiver} =\int Z_{\rm vertex} \, Z_{\rm edge} \ ,
\end{align}
where the integral here could represents either a sum or an integral, or their hybrids.
This ensures that the resulting expression has an interpretation as a statistical mechanical model
with nearest-neighbor as well as self interactions among spins at different sites.

\paragraph{Gauging/Gluing}

Suppose that we have a theory with a flavor symmetry $G$, and write its partition function as $Z_\textrm{before gauging}[a]$,
where $a$ is (at set of) parameter(s) corresponding to the background gauge field for the symmetry $G$.
Let us next gauge the symmetry $G$, to promote it to a gauge symmetry.
The third requirement is that the partition function for the resulting theory takes the form
\begin{align}
Z_\textrm{after gauging}= \int [da] \, Z_\textrm{gauge field}[a] \, Z_{\rm global}[a] \ ,
\label{eq.gauging}
\end{align}
where 
$Z_\textrm{gauge field}[a]$ is the contribution from the
gauge field as well as its superpartner(s), and the integral is over $a$ with appropriate measure $[da]$.
In other words, $Z_\textrm{after gauging}$ with a global symmetry $G$
can be computed in two steps, first keeping the symmetry $G$ as a flavor symmetry and then gauging $G$.

This condition should hold for any consistent localized partition function
(it is a supersymmetric counterpart of the Fubini's theorem for the path integral),
and in many cases (including the 4d $\mathcal{N}=1$ lens index discussed in \cite{Yamazaki:2013nra})
is satisfied trivially. However, there are subtleties for the 2d index, on which we will comment later in this paper.

\paragraph{R-charge}

For the construction of infinite-many conserved charges it is crucial to have 
spectral parameters. In the Gauge/YBE correspondence, R-charge is identified with the R-charges of the bifundamental chiral multiplets. It is therefore important that the partition function 
depends non-trivially on the R-charges of the fields.

\subsection{R-matrix}

Once the conditions above are satisfied, we can write down the integrable model and the R-matrix.
Let us briefly summarize the minimal material, for details readers are referred to \cite{Yamazaki:2013nra}.

The basic idea is to identity the quiver diagram with the lattice of the statistical
mechanical model.

An edge $e\in E$ of the quiver diagram, starting from a vertex $t(e)$ and ending on another $h(e)$,
represents the nearest-neighbor interaction
between the spins at $t(e)$ and $h(e)$.
In gauge theory, this represents a 2d $\mathcal{N}=(2,2)$ chiral multiplet\footnote{2d $\mathcal{N}=(2,2)$ chiral and vector multiplets are dimensional reductions of their 4d $\mathcal{N}=1$ counterparts.} with R-charge $r$, whose
partition function we denote by $\bW_r^e=\bW_r(t(e), h(e))$ ($\bW$ stands for ``weight'').

We have the relation
\begin{align}
\bW_r(a,b) \bW_{2-r}(b,a) =1  \ .
\label{eq.WW=1}
\end{align}
This reflects that fact that the sum of the R-charges of the corresponding chiral multiplets is two,
and hence we can write down a mass term. This means that we can integrate out the fields
in the IR, leading the to trivial partition function as in the right hand side of \eqref{eq.WW=1}.

We also have a 2d $\mathcal{N}=(2,2)$ vector multiplet
at a vertex $v$. This represents the self-interactions among the
spin $s_v$.
We denote the corresponding partition function as
$\bS^v$ ($\bS$ stands for ``self-weight'').

These ingredients, $\bW_r(a,b)$ and $\bS^v$, can be used as the Boltzman weight for the 
definition of the statistical mechanical model, formulate in the language of vertex models:
\begin{align}
Z_{\textrm{2d integrable spin} }= \sum_{\rm spins} \prod_{e: {\rm edge}} \bW^e \prod_{v: {\rm vertex}} \bS^v \ .
\label{ZV}
\end{align}

As in the introduction let us define the model as the Interaction-Round-a-Face (IRF) model,
where the Boltzmann weights are associated with the faces (recall \eqref{eq.YBE}
and \eqref{ZF}).
The partition function for the quiver on the left of
Figure \ref{fig.2dSeiberg_quiver}  is given by
\begin{equation}
\bW_{2-\delta-\alpha}(d,a) \bW_{2-\beta-\gamma}(b,c)
   \sum_g
\bS^g \, \bW_{\alpha}(a,g) \bW_{\beta}(g,b) \bW_{\gamma}(c,g)
\bW_{\delta}(g,d) \ ,
\label{eq.Wdef_orig}
\end{equation}
which by Seiberg-duality coincides with that of the quiver on the right:
\begin{equation}
\bW_{\alpha+\beta}(a,b) \bW_{\gamma+\delta}(c,d)
   \sum_g
\bS^g \, \bW_{1-\alpha}(g,a) \bW_{1-\beta}(b,g) \bW_{1-\gamma}(g,c)
\bW_{1-\delta}(d,g) \ .
\label{eq.Wdef_orig_2}
\end{equation}
The equivalence of these two expressions is the star-star relation \cite{Baxter:1986,Bazhanov:1992jqa}.
Note that we here used the fact that the superpotential in \eqref{W_mag} has R-charge $2$;
it then follows for example that the R-charges for the electric quarks and magnetic quarks sum up to $1$.

The R-matrix, the Boltzmann weight for a face $F$ in the IRF model,
is defined to be a modification of \eqref{eq.Wdef_orig} and \eqref{eq.Wdef_orig}:\footnote{
Equation of (2.9) in version 1 of \cite{Yamazaki:2013nra} contained a typo, which is corrected here.}
\begin{align}
\mathcal{R}(F)&=\matWd{\alpha}{\beta}{\gamma}{\delta}{a}{b}{c}{d} \nonumber \\
&:=
\sqrt{
\frac{\bW_{2-\delta-\alpha}(d,a) \bW_{2-\beta-\gamma}(b,c)}
{\bW_{\alpha+\beta}(a,b)  \bW_{\gamma+\delta}(c,d)}
} \nonumber \\
&\qquad \times 
\sqrt{\bS^a \bS^c}\, \sum_g
\bS^g \, \bW_{\alpha}(a,g) \bW_{\beta}(g,b) \bW_{\gamma}(c,g)
\bW_{\delta}(g,d) \label{eq.Wdef}\\
&=
\sqrt{
\bW_{2-\delta-\alpha}(d,a) \bW_{2-\beta-\gamma}(b,c)
\bW_{2-\alpha-\beta}(b,a) \bW_{2-\gamma-\delta}(d,c)
}  \nonumber \\
&\qquad  \times 
\sqrt{\bS^a \bS^c}\, \sum_g
\bS^g \, \bW_{\alpha}(a,g) \bW_{\beta}(g,b) \bW_{\gamma}(c,g)
\bW_{\delta}(g,d) \ ,
\label{eq.Wdef_2}
\end{align}
where in the last line we used 
the identity \eqref{eq.Wdef_orig}.

For a face $F$,
with external vertices $a, b, c, d$ as in Figure
\ref{fig.YBE},  we denote the rapidity parameters of
four edges by $\alpha, \beta, \gamma, \delta$.
The vanishing of the beta function for the gauge coupling
in the IR
imply that they satisfy the relation
\begin{align}
\alpha+\beta+\gamma+\delta=2 \ .
\label{eq.sum_alpha}
\end{align}

The square root factor of $\sqrt{\bW}$ in \eqref{eq.Wdef}
is the ratio of the prefactors from \eqref{eq.Wdef_orig} and \eqref{eq.Wdef_orig_2};
the square root factor $\sqrt{\bS^a \bS^c}$ is included for the purpose of including the vector multiplet measure into the definition of the R-matrix, so that the sum over the spin in the R-matrix becomes a sum/integral; without this factor we need a non-trivial measure for the sum.

\begin{figure}[htbp]
\centering\includegraphics[scale=0.4]{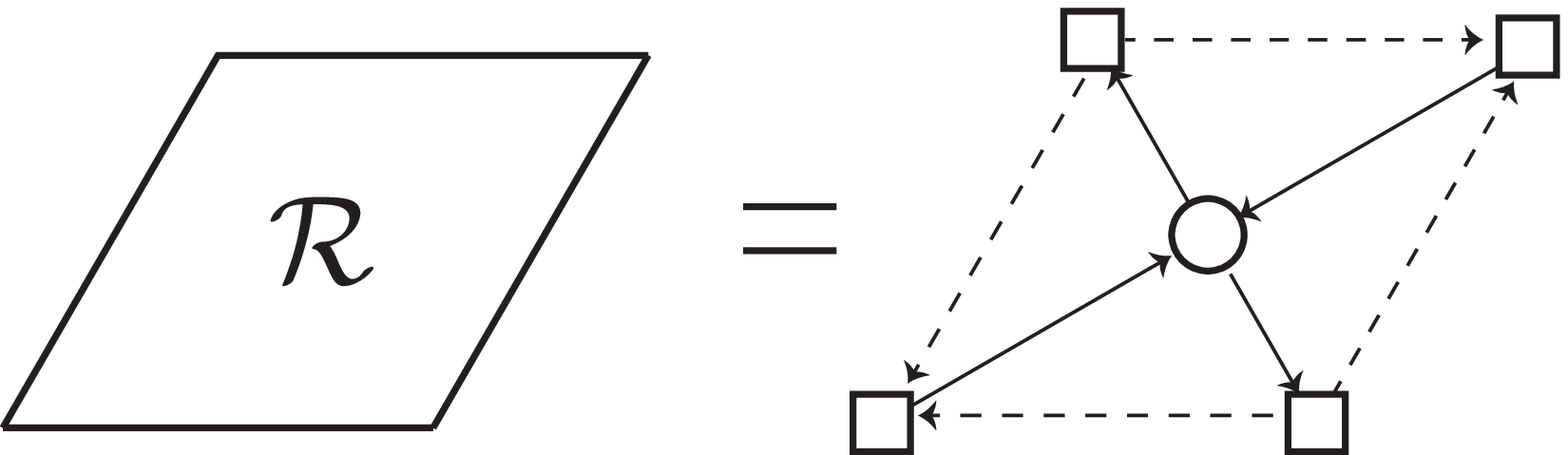}
\label{fig.R_as_quiver}
\caption{The R-matrix is obtained from the partition function of the quiver on the right.}
\end{figure}

\begin{figure}[htbp]
\centering\includegraphics[scale=0.5]{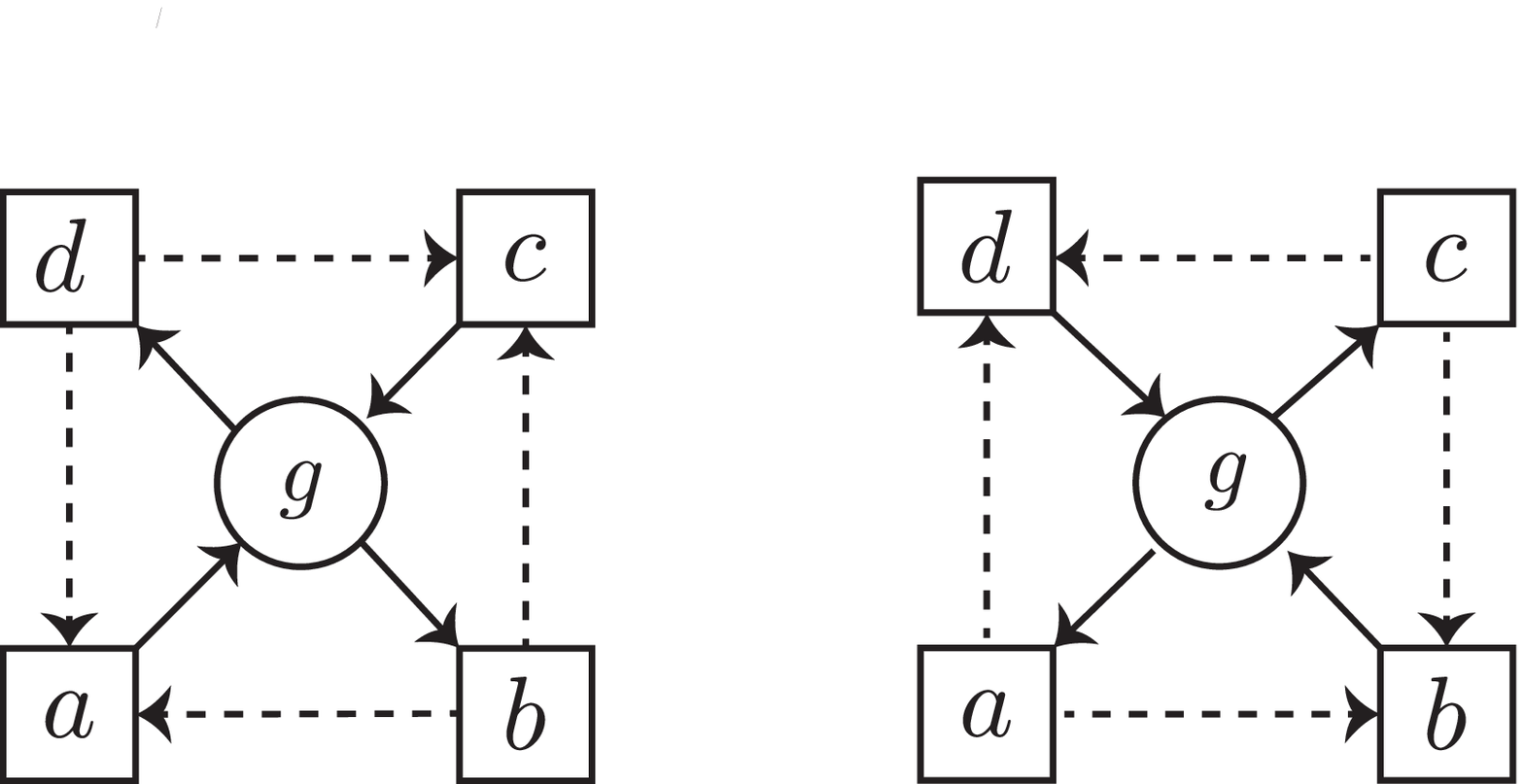}
\label{fig.Seiberg_symmetric}
\caption{By introducing ``half-chiral bifundamental multiplets'' (represented here by a dotted oriented allows),
the Seiberg duality takes a more symmetric form (compare Figure \ref{fig.2dSeiberg_quiver}). The partition function of this quiver coincides with the R-matrix
of the associated integrable model.}
\end{figure}

In terms of gauge theory, this modification is to allow for a ``half-chiral multiplet'', whose partition function is a square root of the 
full (ordinary) chiral multiplet\footnote{
While the notion of a ``half-chiral multiplet'' is natural from the viewpoint of integrable models,
this notion is useful only in so far as the supersymmetric partition functions are discussed, and 
for example there is no such multiplet at the Lagrangian level.
Hence this notion is rather different from the more standard``half-hypermulitiplet'' for 4d $\mathcal{N}=2$ theories. We hope that our nomenclature does not cause any confusion.}. The introduction of a ``half-chiral multiplet'' makes the Seiberg duality more symmetric, as explained in 
Figure \ref{fig.Seiberg_symmetric}. F
In particular, the R-matrix in \eqref{eq.Wdef_2} is nothing but the partition function of the 
quiver of Figure \ref{fig.Seiberg_symmetric}, as represented in Figure \ref{fig.R_as_quiver}:
\begin{align}
\textrm{R-matrix} \longleftrightarrow \textrm{quiver of Figure } \ref{fig.Seiberg_symmetric} \ .
\end{align}

The square root factor in \eqref{eq.Wdef} is a complication, but is necessarily
for the R-matrix to satisfy the YBE in the form of \eqref{eq.YBE}; 
without the prefactor the YBE will be rather the so-called ``twisted YBE''.
This subtlety, which was known in the integrable model literature even before the 
discovery of Seiberg duality, 
is the manifestation of the presence of the mesons in Seiberg duality.
Indeed, the R-matrix as defined in \eqref{eq.Wdef}
has a nice symmetry, thanks to the Seiberg-like duality:
\begin{align}
\matWd{\alpha}{\beta}{\gamma}{\delta}{a}{b}{c}{d} =\sqrt{\frac{\bS_b \bS_d}{ \bS_a \bS_c}} \,  \matWd{1-\alpha}{1-\beta}{1-\gamma}{1-\delta}{b}{c}{d}{a} \ .
\label{eq.R_symmetry}
\end{align}

\begin{figure}[htbp]
\centering\includegraphics[scale=0.55]{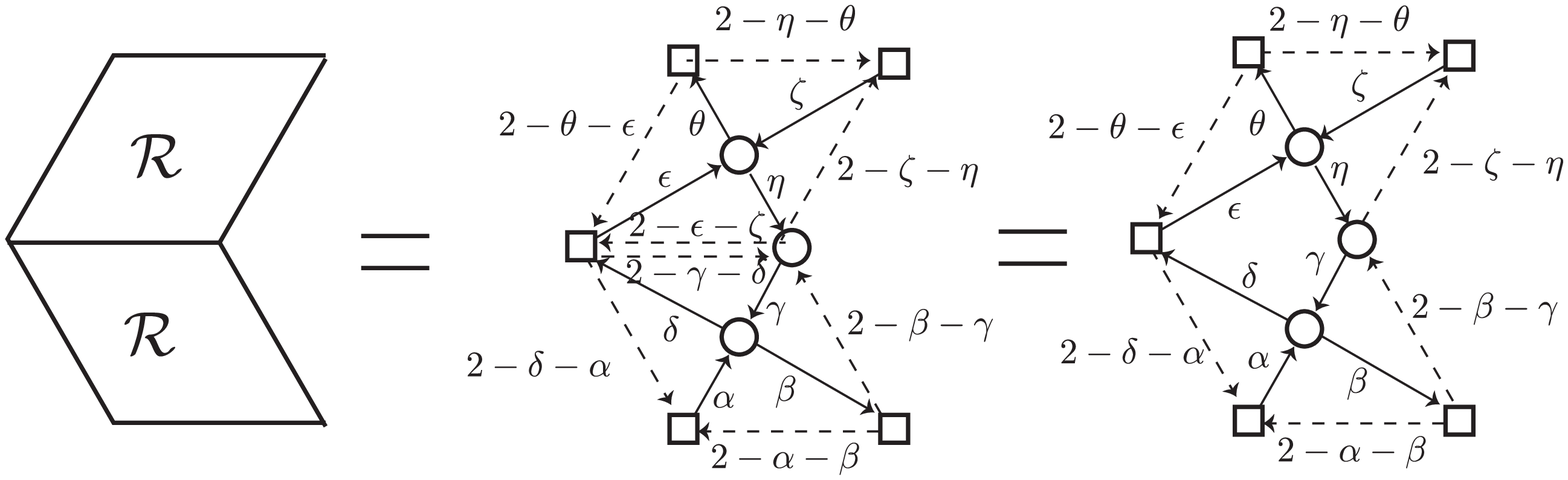}
\caption{Glue two R-matrices corresponds to the concatenation of two quivers by gauging of flavor symmetries. Two ``half-chiral multiplets'' of opposite orientation cancel, while
those with the same orientation combine into an ordinary chiral multiplet.} 
\label{fig.R_glue_YBE_2}
\end{figure}

To discuss YBE, we need to glue the R-matrices.
At the level of the quiver gauge theories, this corresponds to 
concatenating the quiver diagram (see Figure \ref{fig.R_glue_YBE_2}).

When we combine two quivers, it sometimes happens that the half-chiral multiplets are combined together. If they are of the opposite orientations, we cancel them; if of the same orientation, we
replace them by a full chiral multiplet.

To be more precise we have to take into account the constraint from R-charges.
For example, in Figure \ref{fig.R_glue_YBE_2} the two arrows with R-charges 
$2-\epsilon-\zeta$ and $2-\gamma-\delta$ cancels in the end, 
and in order for this to happen we need to be able to write down a mass term for them.
This leads to the constraint
\begin{align}
(2-\epsilon-\zeta) + (2-\gamma-\delta) =2  \ .
\end{align}
This constraint remains even after the chiral multiplet is integrated out,
but then now the equation should lead $\epsilon+\zeta+\gamma+\delta=2$,
namely we generate a new superpotential term corresponding to a face in the middle,
with its superpotential normalized to be $2$.

Following the similar logic, we can verify that the definitions of the partition function
as an IRF model \eqref{ZF} coincides with that as a vertex model \eqref{ZV}.
 
\begin{figure}[htbp]
\centering\includegraphics[scale=0.45]{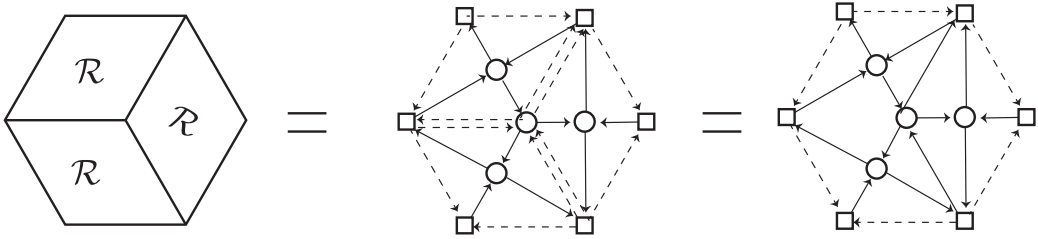}
\caption{We can glue three R-matrices to obtain a quiver corresponding to a product of three R-matrices, the left hand side of \eqref{eq.YBE}.} 
\label{fig.R_glue_YBE}
\end{figure}

\begin{figure}[htbp]
\centering\includegraphics[scale=0.43]{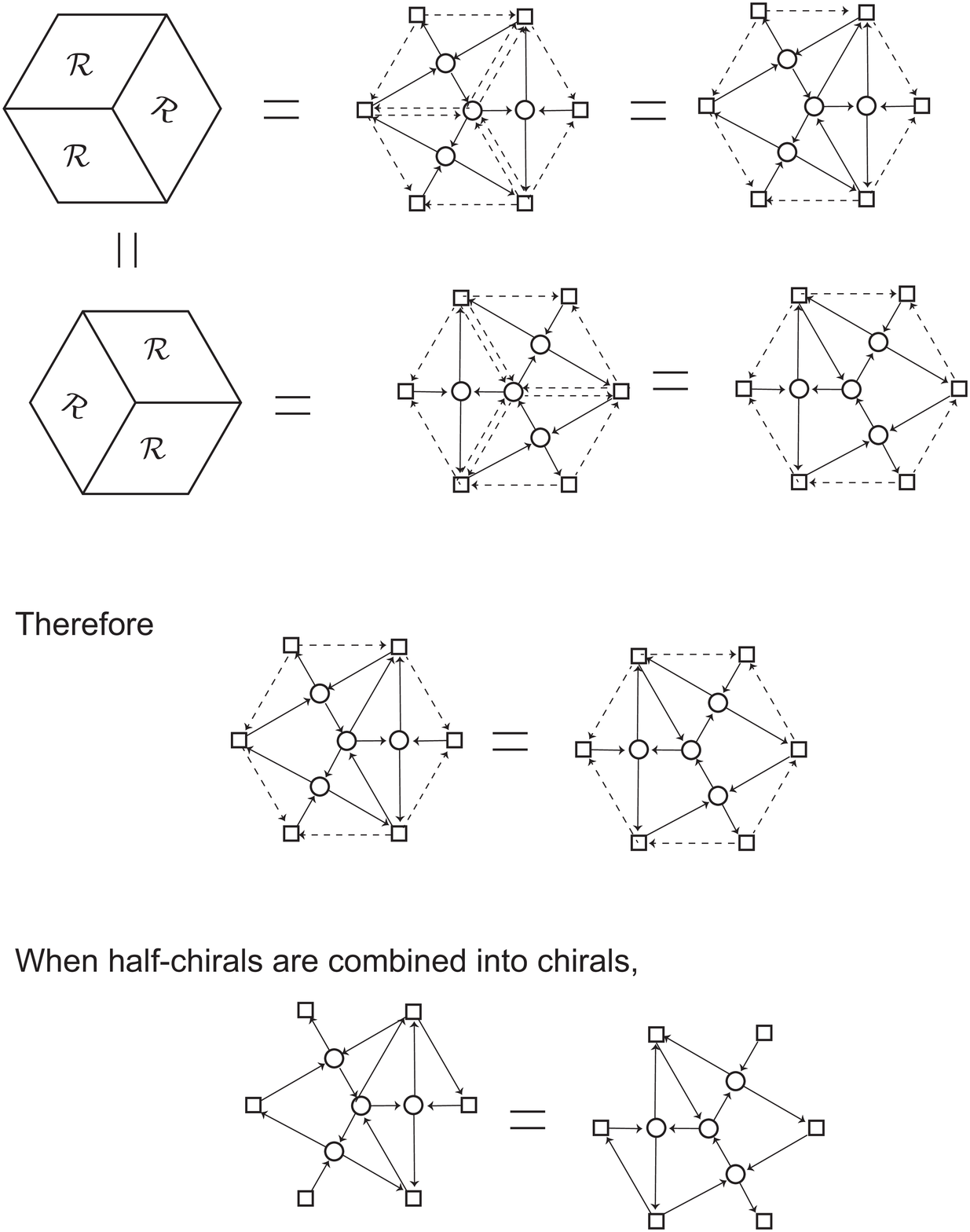}
\caption{For the Yang-Baxter equation, we can glue three R-matrices, following the rule of Figure \ref{fig.R_glue_YBE}. After canceling/combing the half-chiral multiplets
in the equality on both sides of the equation, we obtain the duality of the last line, which was precisely the Yang-Baxter duality of \cite{Yamazaki:2013nra}.} 
\label{fig.R_glue}
\end{figure}

\bigskip
We can repeat this exercise and generate a quiver for the left hand side of 
Figure \eqref{eq.YBE}. A new feature here is that we not have a new internal vertex, which we interprete as a gauge node (see Figure \ref{fig.R_glue_YBE}).
The YBE \eqref{eq.YBE} is now replaced with the Yang-Baxter duality,
a duality between two different quivers
as shown in Figure \ref{fig.R_glue}.

The identity coming from the Yang-Baxter duality then takes the form
\begin{equation}
\begin{split}
&\sum_g
     \left.
\matWd{\alpha}{\beta}{\gamma}{\delta}{a}{b}{g}{f}
\matWd{\iota}{\kappa}{\lambda}{\mu}{g}{b}{c}{d}
\matWd{\epsilon}{\zeta}{\eta}{\theta}{f}{g}{d}{e}
     \, \right|_{\kappa=\gamma+\zeta} \\
&\qquad =
\sum_g
     \left.
\matWd{\epsilon}{\zeta}{\eta}{\theta}{g}{c}{d}{e}
\matWd{\iota}{\kappa}{\lambda}{\mu}{f}{a}{g}{e}
\matWd{\alpha}{\beta}{\gamma}{\delta}{a}{b}{c}{g}
    \, \right|_{\kappa=\gamma+\zeta} 
\ ,
\end{split}
\label{eq:pre_YBE}
\end{equation}
where we have
\begin{align}
\alpha+\beta+\gamma+\delta=\epsilon+\zeta+\eta+\theta=
\iota+\kappa+\lambda+\mu=2 \ ,
\end{align}
as well as the new relation coming from the vanishing of
the beta function for the gauge coupling constant at vertex $g$:
\begin{align}
\kappa=\gamma+\zeta \ .
\end{align}
Since we have total of four constraints for the $4\times 3=12$ parameters, we have
$8$ independent parameters, and these play the role of spectral parameters.

The equation \eqref{eq:pre_YBE}
is already essentially the Yang-Baxter equation \eqref{eq.YBE}.
However the difference is that
Note that our R-matrix $\matWd{\alpha}{\beta}{\gamma}{\delta}{a}{b}{c}{d}$ has three independent spectral parameters
($\alpha, \beta, \gamma, \delta$ with the constraint \eqref{eq.sum_alpha}) ,
while R-matrices $\matWs{u}{a}{b}{c}{d}$ in most of the models in the literature has only one spectral parameter;
the former is more general than the latter.
To specialize to the one-variable R-matrix we choose
\begin{align}
\matWs{u}{a}{b}{c}{d}:=\matWd{\alpha=u}{\beta=1-u}{\gamma=u}{\delta=1-u}{a}{b}{c}{d} \ .
\label{eq.u}
\end{align}
Physically this is situation where $SU(N)^4$ global symmetry (represented by the four $SU(N)$ nodes) enhances to $SU(2N)^2$.
We can then check that \eqref{eq:pre_YBE} reduces to
\eqref{eq.YBE}; the constraint $\kappa=\gamma+\zeta$ accounts for the arguments $u, v, u+v$
in \eqref{eq.YBE}.

In the following we sometimes suppress the spectral parameters and use the shorthand notation $\mathcal{R}\left[\begin{array}{cc}
d & c\\
a & b \\
\end{array}
\right]=\matWd{\alpha}{\beta}{\gamma}{\delta}{a}{b}{c}{d}$.

\section{Integrability from 2d Index}
\label{sec:indexdef}

\subsection{2d Index}
Let us now apply the logic of our previous section to the 
2d index (also known as flavored elliptic genus), namely a supersymmetric partition function
on $T^2$.
For a given 2d $\mathcal{N}=(2,2)$ theory,
the 2d index is defined by
\begin{align}
\mathcal{I}(a_i; q, y) =\textrm{Tr}_{\rm RR} \left[ (-1)^F q^{H_L}\bar{q}^{H_R} y^{J_L} \prod_j a_j^{f_j} \right] \ ,
\end{align}
where the trace is over the RR ground state, where the fermions have periodic boundary conditions.
Also, $F$ is the fermion number, $H_L$ is the left-moving Hamiltonian,
$J_L$ is the left-moving $U(1)$ current and $f_j$'s are the flavor symmetry charges.

The 2d RR index takes the factorized form \cite{Benini:2013nda,Benini:2013xpa}:
\begin{align}
\mathcal{I}_{\rm total}=\int_{\rm Cartan} \mathcal{I}_{\rm chiral} \, \mathcal{I}_{\rm vector} \ ,
\label{T2_integral}
\end{align}
where the integral is over the Cartan of the gauge group.
The index of a single chiral bifundamental multiplet with R-charge $r$
is
\begin{equation}
  \CI_{\rm chiral}(a_i, b_i; q,y)=\prod_{i,j=1}^N\Delta\left(y^{\frac{r}{2}} a_i b_j^{-1};q,y\right) \ ,
\end{equation}
where $a_i, b_i$ denote the Cartan variables for the $SU(N)\times SU(N)$ symmetry of the bifundamental.
The index of a $U(N)$ vector multiplet is
\begin{equation}
  \CI_{\rm vector}^{U(N)}(a_i;q,y)=\frac{1}{N!}\left(\frac{\eta(q)^3}{i\theta_1(y^{-1};q)}\right)^N
  \prod_{1\le i\neq j \le N}\left(\left(1-\frac{a_i}{a_j}\right)\Delta\left(\frac{a_i}{a_j};q,y\right)\right)^{-1} \ .
\end{equation}
Here and in the following we use
\begin{equation}
  \Delta(a;q,y):=\frac{\theta_1(y^{-1}a;q)}{\theta_1(a;q)}  \ ,
\end{equation}
as well as the Dedekind eta function and the Jacobi theta fucntion:
\begin{align}
  &\eta(q):=q^{1/24}\prod_{i=1}^{\infty}(1-q^i) \ ,  \\
  &\theta_1(y;q):=-iq^{\frac{1}{8}}y^{\half}\prod_{i=1}^{\infty}(1-q^i)(1-yq^i)(1-q^{i-1}/y) \ .
\end{align}
For later convenience, 
we also define a shorthand notation of the index of chiral multiplet with R-charge $r$,
\begin{equation}
  \bDelta_r\left(x;q,y\right): =\Delta\left(y^{\frac{r}{2}} x;q,y\right) \ .
\end{equation}
In some cases we omit the arguments $q, y$ to
write $\bDelta_r \left(x\right)$.

\subsection{R-matrix from 2d Index}\label{R_from_2d}
The integrable model has spins
taking values in $z=\{z_i\}=U(1)^{N}$.
This means that the ``sum'' over the spins is actually an integral:
\begin{align}
\sum_v \to \frac{1}{N!} \int_{|z_{v,i}|=1}  \prod_{i=1}^N \frac{dz_{v,i}}{2\pi i z_{v,i}} \ .
\end{align}
The weights are given by
\begin{align}
\bW_e(t(e), h(e))=
\bDelta_{2-r_{h(e)}-r_{t(e)}} \left(z_{t(e),j} z_{h(e),i}^{-1}  ;  q,y\right) 
\end{align}
and
\begin{align}
\bS_v=\frac{1}{N!}\left(\frac{\eta(q)^3}{i\theta_1(y^{-1};q)}\right)^N \prod_{i\neq j}\frac{1}{\Delta\left(z_{v,i} z_{v,j}^{-1};q,y\right)} \ .
\end{align}
The R-matrix is defined by \eqref{eq.Wdef}:
\begin{equation}
\label{eq:U(N)R-matrix}
\begin{split}
  \mathcal{R}&
    \left(\begin{array}{cc}
\delta & \gamma \\
\alpha & \beta \\
\end{array}
\right)
  \left[\begin{array}{cc}
d & c \\
a & b \\
\end{array}
\right](q,y)\\
  =&\sqrt{\frac{ \prod_{i,j=1}^N
        \bDelta_{2-\delta-\alpha}(d_i a^{-1}_j)
        \bDelta_{2-\beta-\gamma}(b_i c_j^{-1})
        }{
        \prod_{i,j=1}^N
        \bDelta_{\alpha+\beta} (a_ib^{-1}_j)
        \bDelta_{\gamma+\delta}(c_id_j^{-1})
        }}
        \left[ \frac{1}{N!}\left(\frac{\eta(q)^3}{i\theta_1(t^{-1};q)}\right)^N \right]^2 \\
    &      \times 
       \sqrt{ \prod_{i\neq j}\frac{1}{\Delta(a_i a_j^{-1};q,y)}  \prod_{i\neq j}\frac{1}{\Delta(c_i c_j^{-1};q,y)}   }
        \oint\prod_{i}\frac{dz_i}{2\pi i z_i}\prod_{i\neq j}\frac{1}{\Delta(z_i z_j^{-1};q,y)}
        \\
        &\times\prod_{i,j=1}^N
        \bDelta_{\alpha}(a_{j} z^{-1}_i)
        \bDelta_{\beta}(z_i b^{-1}_{j})
        \bDelta_{\gamma}(c_{j} z^{-1}_i)
        \bDelta_{\delta}(z_i d^{-1}_{j}) \ .
\end{split}
\end{equation}
This R-matrix satisfies \eqref{eq.R_symmetry} as well as \eqref{eq:pre_YBE}.

There is one subtlety in the present discussion.
To define the answer unambiguously in \eqref{T2_integral} it is important to specify the integration cycle.
This is a rather non-trivial problem, since naively there are poles right on the integration contour.
The general prescription in terms of Jeffrey-Kirwan residue \cite{JeffreyKirwan,SzenesVergne} was given in \cite{Benini:2013xpa},
which unfortunately obscures the factorization \eqref{eq.gauging} (see however
discussion of Appendix \ref{app.2}).

\subsection{Abelian Case}

In Abelian ($U(1)$) case, the expression for R-matrix simplifies dramatically:
\begin{equation}
\begin{split}
  \mathcal{R}^{U(1)}&
\left(\begin{array}{cc}
\delta & \gamma \\
\alpha & \beta \\
\end{array}
\right)
  \left[\begin{array}{cc}
d & c \\
a & b \\
\end{array}
\right](q,y)   
=\sqrt{\frac{ 
        \bDelta_{2-\delta-\alpha}(d a^{-1})
        \bDelta_{2-\beta-\gamma}(b c^{-1})
        }{
        \bDelta_{\alpha+\beta} (a b^{-1})
        \bDelta_{\gamma+\delta}(c d^{-1})
        }}
         \\
    &\qquad      \times 
        \oint\prod_{i}\frac{dz}{2\pi i z} 
        \bDelta_{\alpha}(a z^{-1})
        \bDelta_{\beta}(z b^{-1})
        \bDelta_{\gamma}(c z^{-1})
        \bDelta_{\delta}(z d^{-1}) \ .
\end{split}
\label{U1_int}
\end{equation}
In this case, the Jeffrey-Kirwan residue prescription simplifies, and we can 
appeal to a simpler prescription  
(see \cite{Benini:2013nda,Gadde:2013dda} and Appendix \ref{app.2} for more on this).

To explain this,
let us first note that the position of the poles can be read off from
the identity
\begin{align}
\Delta(z;q,t) = \frac{\theta(t;q)}{(q;q)^2} \sum_{i\in \mathbb{Z}} \frac{t^i}{1-z q^i} \ ,
\end{align}
with poles at $z=q^{-1}$.
Note also that thanks to the relation
\begin{align}
\Delta(q x)=\frac{1}{y} \Delta(x) \ ,
\end{align}
the integrand of \eqref{U1_int} is invariant under the shift $z \to qz$,
and the integrand naturally is a function on a torus.
Due to the residue theorem, we obtain a trivial answer if we combine 
all the residues inside the torus. We should rather pick up a subset of residues,
and the correct choice turns out to pick up those residues with positive (or negative, up to overall 
minus sign of the partition function) charges
\cite{Benini:2013nda,Gadde:2013dda}.
For the case at hand, this amounts to taking the residues at 
$z=a$ and $z=c$, and hence the integral can be worked out explicitly, to obtain
\begin{equation}
\begin{split}
  \mathcal{R}^{U(1)}&
\left(\begin{array}{cc}
\delta & \gamma \\
\alpha & \beta \\
\end{array}
\right)
  \left[\begin{array}{cc}
d & c \\
a & b \\
\end{array}
\right](q,y)     (a,b,c,d; q,y) \\
&=\sqrt{\frac{ 
        \bDelta_{2-\delta-\alpha}(d a^{-1})
        \bDelta_{2-\beta-\gamma}(b c^{-1})
        }{
        \bDelta_{\alpha+\beta} (a b^{-1})
        \bDelta_{\gamma+\delta}(c d^{-1})
        }}
         \\
    &\qquad      \times 
       \left[
        \bDelta_{\alpha+\beta}(a b^{-1})
        \bDelta_{-\alpha+\gamma}(c a^{-1})
        \bDelta_{\alpha+\delta}(a d^{-1})
        + 
        \bDelta_{-\gamma+\alpha}(a c^{-1})
        \bDelta_{\gamma+\beta}(c b^{-1})
        \bDelta_{\gamma+\delta}(c d^{-1})
        \right] \\
           &=
       \sqrt{\frac{
               \bDelta_{\alpha+\delta}(a d^{-1})
         \bDelta_{\alpha+\beta}(a b^{-1})
        }{
        \bDelta_{\gamma+\beta}(c b^{-1})
        \bDelta_{\gamma+\delta}(c d^{-1})
        }}
        \bDelta_{\gamma-\alpha}(c a^{-1})
        +
        \sqrt{\frac{ 
        \bDelta_{\gamma+\beta}(c b^{-1})
        \bDelta_{\gamma+\delta}(c d^{-1})
        }{
        \bDelta_{\alpha+\beta} (a b^{-1})
        \bDelta_{\alpha+\delta}(a d^{-1})
        }} 
  \bDelta_{\alpha-\gamma}(a c^{-1})
        \ .
\end{split}
\end{equation}

\subsection{Modularity}

The RR index on $T^2$ has a natural modular property. Under the modular transformation, the Dedekind eta function and Jacobi theta function transform as follows: 
\begin{align}
  &\eta\left(e^{2\pi i(-\frac{1}{\tau})}\right)=\sqrt{-i\tau}\, \eta(e^{2\pi i\tau}) \ ,\\
  &\theta_1\left(e^{2\pi i(-\frac{\zeta}{\tau})};e^{2\pi i(-\frac{1}{\tau})}\right)
   =i\sqrt{-i\tau}\, e^{\pi i\frac{\zeta^2}{\tau}}\, \theta_1\left(e^{2\pi i \zeta};e^{2\pi i\tau}\right) \ ,
\end{align}
where $q:= e^{2\pi i\tau}$ and $y:= e^{2\pi i\zeta}$ are the original parameters, while $\tq: =e^{2\pi i(-\frac{1}{\tau})}$ and $\ty: =e^{2\pi i(-\frac{\zeta}{\tau})}$ are the parameters after modular transformation.

The modular property of a chiral multiplet with R-charge $r$ and some flavor fugacity $a$ is then
\begin{equation}
  \begin{split}
    \Delta(\ty^{\frac{r}{2}}\tilde{a};\tq,\ty)
    &=\frac{\theta_1(e^{2\pi i(\frac{r}{2}-1)(-\frac{\zeta}{\tau})}e^{2\pi i(-\frac{\upsilon_a}{\tau})};e^{2\pi i(-\frac{1}{\tau})})}
    {\theta_1(e^{2\pi i\frac{r}{2}}(-\frac{\zeta}{\tau})e^{2\pi i(-\frac{\upsilon_a}{\tau})};e^{2\pi i(-\frac{1}{\tau})})}\\
    &=e^{\frac{\pi i}{\tau}[(1-r)\zeta^2-2\upsilon_a\zeta]}
     \frac{\theta_1(e^{2\pi i(\frac{r}{2}-1)\zeta}e^{2\pi i\upsilon_a};e^{2\pi i\tau})}
    {\theta_1(e^{2\pi i\frac{r}{2}\zeta}e^{2\pi i\upsilon_a};e^{2\pi i\tau})}\\
    &=e^{\frac{\pi i}{\tau}[(1-r)\zeta^2-2\upsilon_a\zeta]}\Delta(y^{\frac{r}{2}}a;q,t) \ ,
  \end{split}
\end{equation}
where we defined $\upsilon_a$ by $a=:e^{2\pi i\upsilon_a}$. Notice that the coefficient of $\zeta^2$ term in the modular weight is $1/3$ of the central charge of the chiral multiplet,
\begin{equation}
  c=3\,\tr\gamma_3J^2_L=3\left(\left(\frac{r}{2}-1\right)^2-\left(\frac{r}{2}\right)^2\right)=3(1-r) \ ,
\end{equation}
and the coefficient of linear term  is $-\CA^a\upsilon_a$ where $\CA^a$ is the anomaly of flavor symmetry $F_a$,
\begin{equation}
  \CA^a=\tr\gamma_3J_LF_a=\left(\left(\frac{r}{2}-1\right)-\left(\frac{r}{2}\right)\right)F_a \ .
\end{equation}
Similarly, under modular transformation the vector multiplet 
$\mathcal{I}^{U(N)}_{\rm vector}(z_i;q,t)$ behaves as
\begin{equation}
  \left(\frac{\eta(\tq)^3}{i\theta_1(\ty^{-1};\tq)}\right)^N
  \prod_{i\neq j}\Delta\left(\frac{\tilde{z}_i}{\tilde{z}_j};\tq,\ty\right)^{-1}
  =(-\tau)^N e^{-\pi i N^2\frac{\zeta^2}{\tau}}\left(\frac{\eta(q)^3}{i\theta_1(y^{-1};q)}\right)^N
  \prod_{i\neq j}\Delta\left(\frac{z_i}{z_j};q,y\right)^{-1}.
\end{equation}
Again the coefficient of $\zeta^2$ gives the correct central charge for vector multiplets. In general, one can read off the central charge and anomalies for a theory from the modular weight of its index,
\begin{equation}
  \CI(e^{2\pi i(-\frac{\upsilon_a}{\tau})};e^{2\pi i(-\frac{1}{\tau})},e^{2\pi i(-\frac{\zeta}{\tau})})
  =e^{\frac{i\pi}{\tau}\left(\frac{c}{3}\zeta^2-2\CA^a \upsilon_a\zeta\right)}\CI(e^{2\pi i\upsilon_a};e^{2\pi i\tau},e^{2\pi i \zeta}) \ .
\end{equation}
One can then derive the modular property of R-matrix \eqref{eq:U(N)R-matrix},
\begin{equation}
\label{eq:modularU(N)}
 \mathcal{R}\left(\begin{array}{cc}
\delta & \gamma \\
\alpha & \beta \\
\end{array}
\right)\!
  \left[\begin{array}{cc}
\tilde{d} & \tilde{c} \\
\tilde{a} & \tilde{b} \\
\end{array}
\right]
  =(-\tau)^N e^{\frac{i\pi}{\tau}\frac{3N^2}{3}\zeta^2}
   \mathcal{R}\left(\begin{array}{cc}
\delta & \gamma \\
\alpha & \beta \\
\end{array}
\right) \!
  \left[\begin{array}{cc}
d & c \\
a & b \\
\end{array}
\right]
    \ .
\end{equation}
The number $3N^2$ represents the central charge $3 N^2$,
which can be directly computed from the spectrum of the theory\footnote{
There is an analogous statement for the 4d $\mathcal{N}=1$ quiver gauge theories:
the high temperature limit of the $S^1\times S^3$ is reproduces
a linear combination of central charges $a, c$ \cite{DiPietro:2014bca}.
}.

\subsection{Dimensional Reduction}

We can consider the dimensional reduction of our 2d $\mathcal{N}=(2,2)$ theory on
$S^1$. We expect that our 2d $\mathcal{N}=(2,2)$ theory will reduce to
1d $\mathcal{N}=4$ supersymmetric quantum mechanics, and the 2d Seiberg-like duality to a
duality in 1d.

In our setup, we can take choose the $S^1$ of the dimensional reduction to be one of the cycles of $T^2$.
This is the reduction to the 1d index derived recently in \, \cite{Hori:2014tda,Cordova:2014oxa}.
In Appendix we also derive the same result by the reduction procedure in Appendix \ref{sec:2d1d}.

The quantity relevant for us is the Witten index for the $\mathcal{N}=4$ supersymmetric quantum mechanics,
twisted by a subgroup of the R-symmetry commuting with the supercharge:
\begin{align}
\textrm{Tr} \left[
(-1)^F e^{2\pi i z J_-} e^{-\beta H}
\right] \ .
\end{align}
When we regard $\mathcal{N}=4$ quantum mechanics as $\mathcal{N}=2$ quantum mechanics,
part of the $\mathcal{N}=4$ R-symmetry looks like a flavor symmetry for the $\mathcal{N}=2$ theory,
and $J_{-}$ generates the flavor symmetry there.

The Boltzmann weight for an edge, i.e.\ the 1-loop determinant for a
bifundamental chiral multiplet is\footnote{In the literature, 
the function $\sinh$ is replaced by $\sin$ in \eqref{W_1d} and \eqref{S_1d}.
This amounts to the rotation of the contour,
which does not affect the answer as long as we sum over the same set of residues.
The expression with $\sinh$ is the one which naturally arises from the dimensional reduction, see appendix \ref{sec:2d1d}.}
\begin{equation}
\bW^e(\bm{a}_{t(e)}, \bm{a}_{h(e)})= \prod_{i,j=1}^N  \frac{\sinh(\bm{a}_{t(e), i}-\bm{a}_{h(e), j}+z)}{\sinh(\bm{a}_{t(e), i}-\bm{a}_{h(e), j})} \ ,
\label{W_1d}
\end{equation}
where the 2d integrable variables $a_i$ are here replaced by their 1d counterparts $\bm{a}_i$.
The Boltzmann weight for a vertex, i.e.\ the 1-loop determinant for a
vector multiplet is
\begin{equation}
\bS^v(\bm{a}_v)= \prod_{i\ne j}  \frac{\sinh(\bm{a}_{v, i}-\bm{a}_{v, j} )}{\sinh(\bm{a}_{v, i}-\bm{a}_{v, j}+z)} \ .
\label{S_1d}
\end{equation}

The R-matrix is given by
\begin{align}
\begin{split}
 &\mathcal{R}
    \left(\begin{array}{cc}
\delta & \gamma \\
\alpha & \beta \\
\end{array}
\right)
  \left[\begin{array}{cc}
\bm{d} & \bm{c} \\
\bm{a} & \bm{b} \\
\end{array}
\right](z)\\
  =&\sqrt{\frac{ \prod_{i,j=1}^N
        \bbDelta_{2-\delta-\alpha}(\bm{d}_i-\bm{a}_j)
        \bbDelta_{2-\beta-\gamma}(\bm{b}_i-\bm{c}_j)
        }{
        \prod_{i,j=1}^N
        \bbDelta_{\alpha+\beta} (\bm{a}_i-\bm{b}_j)
        \bbDelta_{\gamma+\delta}(\bm{c}_i-\bm{d}_j)
        }}
         \\
         &\times \frac{1}{N!^2} \sqrt{\prod_{i\neq j}\frac{1}{\bm{\Delta}(\bm{a}_i-  \bm{a}_j,z)}\prod_{i\neq j}\frac{1}{\bm{\Delta}(\bm{c}_i - \bm{c}_j,z)}}
        \oint\prod_{i} d\bm{z}_i\prod_{i\neq j}\frac{1}{\bm{\Delta}(\bm{z}_i -\bm{z}_j,z)}
        \\
        &\times\prod_{i,j=1}^N
        \bbDelta_{\alpha}(\bm{a}_{j}-\bm{z}_i)
        \bbDelta_{\beta}(\bm{z}_i - \bm{b}_{j})
        \bbDelta_{\gamma}(\bm{c}_{j}-  \bm{z}_i)
        \bbDelta_{\delta}(\bm{z}_i - \bm{d}_{j}) \ ,
\end{split}
\end{align}
where we defined
\begin{align}
\bm{\Delta}(\bm{x}, z)& := \frac{\sinh\left(\bm{x}-z\right)}{\sinh\left(\bm{x}\right)} \ , \\
\overline{\bm{\Delta}}_r(\bm{x}, z)& := \frac{\sinh\left(\bm{x}+(\frac{r}{2}-1)z\right)}{\sinh\left(\bm{x}+\frac{r}{2}z\right)} \ .
\end{align}
The integral is again understood to be defined in terms of the Jeffrey-Kirwan residue.
For $U(1)$ theories, the contour prescription gives
\begin{equation}
  \begin{split}
   \mathcal{R}^{U(1)}&
    \left(\begin{array}{cc}
\delta & \gamma \\
\alpha & \beta \\
\end{array}
\right)
  \left[\begin{array}{cc}
\bm{d} & \bm{c} \\
\bm{a} & \bm{b} \\
\end{array}
\right](z)\\
    =&
 \sqrt{
\frac{   \bbDelta_{\alpha+\delta}(\bm{a}-\bm{d}, z)
 \bbDelta_{\alpha+\beta}(\bm{a}-\bm{b}, z)
  }
   {
    \bbDelta_{\gamma+\beta}(\bm{c}-\bm{b}, z)
   \bbDelta_{\gamma+\delta}(\bm{c}-\bm{d}, z)
   }
   }
   \bbDelta_{\gamma-\alpha}(\bm{c}-\bm{a}, z)
      \\
 &+
 \sqrt{
\frac{   \bbDelta_{\gamma+\beta}(\bm{c}-\bm{b}, z)
 \bbDelta_{\gamma+\delta}(\bm{c}-\bm{d}, z)
  }
   {
    \bbDelta_{\alpha+\beta}(\bm{a}-\bm{b}, z)
   \bbDelta_{\alpha+\delta}(\bm{a}-\bm{d} ,z)
   }
   }
   \bbDelta_{\alpha-\gamma}(\bm{a}-\bm{c}, z)
\ .
  \end{split}
\end{equation}

\section{Comments on Brane Realizations}\label{sec.brane_realization}

One novel aspect of the Gauge/YBE correspondence is that
the integrability resides not in each individual quiver gauge theory, but in a {\it class} of gauge theories. In other words, integrability is in the ``theory space''\footnote{
The exploration of any structure of the theory space is a fascinating topic.
For recent attempts, see e.g.\ entanglement \cite{Yamazaki:2013xva} and cluster algebras \cite{Benini:2014mia,Terashima:2013fg}.
}, and to properly 
understand for example the meaning of conserved charged in integrable models
we are required to go beyond the familiar territory of conventional quantum field theories,
and discuss the theory space inside a new framework.

One candidate for such a framework is the string theory---it is expected
that different gauge theories are realized as different configuration of branes,
and branes themselves should be regarded as dynamical degrees of freedom in the 
string theory. It is therefore natural to discuss the string theory realizations of the quiver gauge theories, as a hint for the existence of integrable structure therein.

\bigskip

Let us here study the case of 4d $\mathcal{N}=1$ quiver gauge theories discussed in \cite{Yamazaki:2012cp,Terashima:2012cx,Yamazaki:2013nra};
we can realize 2d $\mathcal{N}=(2,2)$ theories by dimensional reduction (or T-duality) of these theories.
For this case, the relevant brane configurations are known,
both for torus quivers and planar quivers.

Let us for concreteness consider case of the torus quivers.
The relevant type IIB brane configuration \cite{Imamura:2006ie,Imamura:2007dc,Yamazaki:2008bt} (mirror to type IIA description of \cite{Feng:2005gw}) is shown in Table \ref{tbl:D5NS5}.

In this brane configuration,  we consider type IIB string theory on 
$\mathbb{R}^{3,1}\times (\mathbb{C}^{\times})^2\times \mathbb{C}$.
$N$ D5-brane spread in $012357$-directions,
with $57$ directions compactified (namely gives $T^2$). We in addition have one NS5-brane, 
filling the $0123$ directions as well as 
a holomorphic curve 
\begin{align}
\Sigma(x,y)=0  \ ,  \quad x=e^{x_4+i x_5} \ , \quad y=e^{x_6+i x_7}  \ ,
\end{align}
inside $(\mathbb{C}^{\times})^2$ ($4567$-directions).
The NS5-brane and D5-brane intersect along 1-cycles, which divides the $T^2$
into various regions, realizing the quiver structure of the gauge group.

\begin{table}[htpb]
\begin{center}
\caption{The five-brane configuration realizing 4d $\mathcal{N}=1$ quiver gauge theories.}
\begin{tabular}{c|cccc|cccc|cc}
\hline
\hline
&0&1&2&3&4&5&6&7&8&9 \\
\hline
D5&$\circ$ &$\circ$ &$\circ$ &$\circ$ & & $\circ$ & &$\circ$& & \\
NS5&$\circ$ &$\circ$ &$\circ$ &$\circ$ &\multicolumn{4}{c}{$\Sigma$ (2-dim surface)} & \multicolumn{1}{|c}{} \\
\hline
\end{tabular}
\label{tbl:D5NS5}
\end{center}
\end{table}

Now the crucial point here is the Seiberg duality can be understood as a rearrangement of these 1-cycles; in fact, as pointed out in \cite{Hanany:2005ss,Yamazaki:2012cp}
the rapidity lines of integrable models match precisely with the zig-zag paths, which in the language of \cite{Imamura:2006ie,Imamura:2007dc,Yamazaki:2008bt} are the intersection of D5-branes with NS5-branes, and YBE is to exchange the relative position of these 1-cycles\footnote{
More precisely, YBE is really a double Yang-Baxter move\cite{Yamazaki:2012cp}, namely YBE applied twice.}.

What do we gain from this? The basic idea is that we should then be able to describe integrability
in terms of the effective theory on the 5-branes (say $N$ D5-branes). The intersection with NS5-brane in this viewpoint 
appears as BPS defects inside that theory, and consequently the integrability is translated into the 
statements about the rearrangement of BPS defects.
This way, the problem of ``theory space'' is turned into a more tractable problem of 
the discussion of BPS defects inside a supersymmetric gauge theory. We still obtain a class of theories in the sense that the insertion of BPS defects (disorder operators) change the definition of the path integral, however at least the starting point is always a single gauge theory.

The idea that integrability follows from rearrangement of defects, especially line defects,
goes back to the discussion of line defects in pure Chern-Simons theory \cite{Witten:1989wf},
and more recently in the work of \cite{Costello:2013zra}. The latter reference in particular
realizes the spectral parameters, which are absent in the descriptions of \cite{Witten:1989wf}.

These considerations naturally lead us to the question if there are any relations between the present work and the work of \cite{Costello:2013zra}. While there are many similarities,
one cautionary remark is that the actual integrable models studied in \cite{Costello:2013zra}
is the Heisenberg XXX spin chain model, whereas the integrable models discussed in \cite{Yamazaki:2013nra} and here are tend to be more complicated models (chiral Potts models and six-vertex models, and their sophisticaed generalizations), and 
while there are some connections \cite{Bazhanov:1989nc} the connection is at best not direct.

Despite this cautionary remark, the similarity is fascinating, and it would be an interesting problem to pursue this type of reasoning futher, to elucidate the integrable structures in the Gauge/YBE correspondence. One possible clue is that, after the compactification of $3$-direction, the 5-brane systems of Table \ref{tbl:D5NS5}
is dual to the description of codimension 2 defects inside the M5-brane theory.

\section{Conclusion}\label{sec.conclusion}

In this paper we constructed integrable models (solutions to the YBE)
from the $T^2$ partition function of the 
2d $\mathcal{N}=(2,2)$ quiver gauge theories and the dualities
among them (namely Seiberg-like duality and Yang-Baxter duality).

The resulting integrable model has an R-matrix written in term of theta functions,
and the former has a nice modularity behavior.
As an example of the reduction, we worked out the reduction of the 2d index to the 1d index of the 
dimensionally-reduced $\mathcal{N}=4$ theory.

Along the way we clarified some technical aspects of the Gauge/YBE correspondence,
and also encountered several new ingredieents, which are not present in their
4d counterparts \cite{Yamazaki:2013nra}.

Here are some open questions:
\begin{itemize}

\item It would be interesting to compare our solutions to the known solutions in the literature,
and also the identify the quantum-group-like structure underlying our solutions.
There are well-known solutions of YBE in terms of theta function, however
our model are atypical in that we have continuous 
spin variables.

\item Given a solution of the integrable model, we can study its degeneration; the integral model of \cite{Yamazaki:2013nra} reproduce in this way many known integrable models, including the Ising models and their generalizations. In particular, the root of unity 
degeneration of the model gave rise to integrable models with discrete spins.
It would hence be interesting to study the root-of-unity degeneration of our models.

\item We can try to replace the $T^2$ partition function by the $S^2$ partition function \cite{Benini:2012ui,Doroud:2012xw}. One subtlety in this case is that 
the $S^2$ partition function depends on the complexified FI parameter,
which transforms non-trivially under the Seiberg-like duality. In fact,
it was observed in \cite{Benini:2014mia} that the FI parameter transforms as a cluster $y$-variable in the theory of cluster algebras \cite{FominZelevinsky4}. This means that the statistical model coming from the $S^2$ partition function does not 
solve the standard YBE, but rather a generalization of it, where integrability is combined with the cluster algebra.

\item The vacua of 2d $\mathcal{N}=(2,2)$ theories are described by Bethe Ansatz equation of integrable model (Gauge/Bethe correspondence \cite{Nekrasov:2009uh}).
It would be interesting to understand the relation between Gauge/Bethe correspondence and Gauge/YBE correspondence. Let us here point out that the integrable structure there is of a rather different nature from
the integrable structured discussed in this paper. First,
the precise integrable models there are XXX models and their generalizations, while here
we have the six-vertex models and their cousins.
Second, in \cite{Nekrasov:2009uh} the Bethe Ansatz equations play a role, whereas here 
the Boltzmann weights and the R-matrix of the integrable models play direct roles.
Third, \cite{Nekrasov:2009uh} is about the vacuum structure of 2d $\mathcal{N}=(2,2)$ theories, whereas
our story here is about the supersymmetric partition functions of the 2d $\mathcal{N}=(2,2)$ theories.

\item 
 It is natural to consider the  Gauge/YBE correspondence for the 4d $\mathcal{N}=1$ partition function on $S^2\times T^2$ \cite{Closset:2013sxa,Benini:2015noa,Honda:2015yha}.
Unfortunately, the R-charge on $S^2\times T^2$ should satisfy an integrability constraint
and consequently continuous spectral parameter seems to be lost in this setup.

\item 
The 4d $\mathcal{N}=1$ theories discussed in \cite{Yamazaki:2013nra}
is known to have explicit brane realizations (see e.g.\ \cite{Yamazaki:2008bt,Heckman:2012jh}, as well as section \ref{sec.brane_realization}).
Naively one might imagine that we can dimensionally reduce these 4d $\mathcal{N}=1$ theories on $T^2$, to obtain the 2d $\mathcal{N}=2$ theories discussed here, leading to the $T^2$ compactification of relevant brane configurations. However the dimensional reduction of the 4d Seiberg duality requires a careful analysis \cite{Aharony:2013dha}, in particular for quiver gauge theories, and it remains to be carry out these reduction in detail on the field theory side.
This type of analysis is need also for the proper field theory understanding of the 3d $\mathcal{N}=2$ theories discussed in \cite{Yamazaki:2012cp,Terashima:2012cx}.

\item The Yang-Baxter equation has a higher dimensional generalization, for example in three dimensions the relevant equation is the tetrahedron equation. The question is if these equations has their supersymmetric counterparts (see \cite{GaddeYamazaki} for a recent result).

\end{itemize}

\section*{Acknowledgements}
We would like to thank Simons Center for geometry and physics and the 2014 Simons workshop for hospitality, where this work was initiated.
We would like to thank Ibou Bah, Jacque Distler, Abhijt Gadde, Kentaro Hori, Ken Intriligator, Bei Jia, Andy Neitzke, Jaewon Song and Cumrun Vafa  for stimulating discussion.
The contents of this talk was presented by WY at USC (Oct. 2014), UT Austin (Nov. 2014), Caltech (Dec. 2014), and UCSD (Feb. 2015), and by MY at IPMU (Nov. 2014) and KIAS (Dec. 2014).
The work of WY is supported in part by the Sherman Fairchild scholarship, by DOE grant DE-FG02-92- ER40701, and by Walter Burke Institute for theoretical physics.
The research of MY is supported in part by the World Premier International Research Center
Initiative (MEXT, Japan), by  JSPS Program for Advancing Strategic
International Networks to Accelerate the Circulation of Talented Researchers,
by JSPS KAKENHI Grant Number 15K17634,
and by Institute for Advanced Study.

\appendix

\section{Proof of YBE for $T^2$ Partition Function}\label{app.2}

In this appendix, we explicitly prove the YBE for the $T^2$ partition function.
We hope this is useful for the mathematical-oriented readers who wish to skip the 
gauge theories dualities, and illustrate our contour prescription, which is crucial for the 
YBE.

\subsection{Proof of 2d Seiberg-like Duality for 2d Index}

Let us write down the 2d index for the right quiver in Figure \ref{fig.2dSeiberg_quiver}:
\begin{equation}
\begin{split}
&  \CI_{\{r_a,r_b,r_c,r_d\}}(a,b,c,d; q,y)\\ &\qquad=\frac{1}{N!}\left(\frac{\eta(q)^3}{i\theta_1(y^{-1};q)}\right)^N 
        \prod_{i,j=1}^N
        \bDelta_{2-r_a-r_b}(a_ib^{-1}_j)
        \bDelta_{2-r_c-r_d}(c_id_j^{-1})\\\
       &\qquad \times  \oint\prod_{i}\frac{dz_i}{2\pi i z_i}\prod_{i\neq j}\frac{1}{\Delta(z_i z_j^{-1};q,y)}
        \\
        &\qquad \times\prod_{i,j=1}^N
        \bDelta_{r_a}(z_{j} a^{-1}_i)
        \bDelta_{r_b}(b_i z^{-1}_{j})
        \bDelta_{r_c}(z_{j} c^{-1}_i)
        \bDelta_{r_d}(d_i z^{-1}_{j}) \ .
\end{split}
\end{equation}
The fundamental identity representing the 2d Seiberg-like duality is
\begin{equation}
\label{eq:fundSD}
  \CI_{\{r_a,r_b,r_c,r_d\}}(a,b,c,d; q,y)=\CI_{\{1-r_b,1-r_c,1-r_d,1-r_a\}}(b,c,d,a;q,y) \ .
\end{equation}
We first prove \eqref{eq:fundSD}.

Let us define the following index,
\begin{equation}
  \CI_{\rm SD}(\bfA,\bfB;q,y)
 :  =\frac{1}{N!}\left(\frac{\eta(q)^3}{i\theta_1(y^{-1};q)}\right)^N \!
   \oint\!\prod_{\alpha=1}^N\frac{dz_i}{2\pi i z_\alpha}
   \frac{\prod_{\alpha=1}^N\prod_{i=1}^{2N}\Delta(z_\alpha A_i^{-1};q,y)\Delta(B_iz^{-1}_\alpha;q,y)}
   {\prod_{\alpha\neq\beta}\Delta(z_\alpha z_\beta^{-1};q,y)} \ ,
\end{equation}
which can be viewed as the index of a 2d $(2,2)$ theory with $U(N)$ gauge group and $2N$ fundamental and anti-fundamental chirals. $\bfA$ and $\bfB$ are shorthand notation for $\{A_i\}$ and $\{B_i\}$, and we set
\begin{equation}
  \label{app:eq:polestructure}
  \begin{split}
    &|A_i|>1,\quad\quad|qA_i|<1,\quad\quad i=1,\cdots,2N \ ,\\
    &|B_i|<1,\quad\quad|q^{-1}B_i|>1,\quad\quad i=1,\cdots,2N \ .
  \end{split}
\end{equation}

Now let us prove the following identity (this is already in version 2 in \cite{Gadde:2013dda}):
\begin{equation}
  \label{app:eq:fundSD}
  \CI_{\rm SD}(\bfA,\bfB;q,y)=\prod_{i,j=1}^{2N}\Delta(B_j/A_i;q,y)\CI_{\rm SD}(y^{-\half}\bfA^{-1},y^{\half}\bfB^{-1};q,y) \ ,
\end{equation}
where $\bfA^{-1}$ and $\bfB^{-1}$ are shorthand notation for $\{A^{-1}_i\}$ and $\{B^{-1}_i\}$.
The left-hand-side of \eqref{app:eq:fundSD} can be computed by the residue prescription in \cite{Gadde:2013dda}. The poles can be picked as
\begin{equation}
  z_\alpha=B_{i_\alpha} \ ,
\end{equation}
where all the other poles in side the unit circle like $q^kA_i$'s and $q^kB_i$'s have opposite residues and will not contribute to the integral. The result is
\begin{equation}
  \begin{split}
    \CI_{\rm SD}(\bfA,\bfB;q,y)
    &=\sum_{\{i_\alpha\}}\prod_{\alpha=1}^N\prod_{j=1}^{2N}\Delta(B_{i_\alpha} A_j^{-1};q,y)
     \frac{\prod_{\alpha=1}^N\prod_{j\neq i_\alpha}^{2N}\Delta(B_j B_{i_\alpha}^{-1};q,y)}{\prod_{\alpha\neq\beta}\Delta(B_{i_\alpha} B_{i_\beta}^{-1};q,y)}\\
    &=\sum_{\{i_\alpha\}}\prod_{s\in\{i_\alpha\}}\prod_{j=1}^{2N}\Delta(B_{i_\alpha} A_j^{-1};q,y)
      \prod_{s\in\{i_\alpha\}}\prod_{r\in\overline{\{i_\alpha\}}}\Delta(B_r B_s^{-1};q,y) \ .
  \end{split}
\end{equation}
The right-hand-side of equation \ref{app:eq:fundSD} can be computed similarly. We pick up the residues for the poles at
\begin{equation}
  z_\alpha=y^{\half} B^{-1}_{i_\alpha} \ ,
\end{equation}
and the result is
\begin{equation}
  \begin{split}
     &\CI_{\rm SD}(y^{-\half}\bfA^{-1},y^{\half}\bfB^{-1};q,t)\\
    &\qquad=\sum_{\{i_\alpha\}}\prod_{\alpha=1}^N\prod_{j=1}^{2N}\Delta(y A_j B_{i_\alpha}^{-1};q,y)
      \frac{\prod_{\alpha=1}^N\prod_{j\neq i_\alpha}^{2N}\Delta(B_{i_\alpha} B_j^{-1};q,y)}{\prod_{\alpha\neq\beta}\Delta(B_{i_\beta} B_{i_\alpha}^{-1};q,y)}\\
    &\qquad=\sum_{\{i_\alpha\}}\prod_{s\in\{i_\alpha\}}\prod_{j=1}^{2N}\Delta(y A_j B_s^{-1};q,y)
      \prod_{s\in\{i_\alpha\}}\prod_{r\in\overline{\{i_\alpha\}}}\Delta(B_s B_r^{-1};q,y)\\
    &\qquad=\sum_{\overline{\{i_\alpha\}}}\prod_{s\in\overline{\{i_\alpha\}}}\prod_{j=1}^{2N}\Delta(y A_j B_s^{-1};q,y)
      \prod_{s\in\overline{\{i_\alpha\}}}\prod_{r\in\{i_\alpha\}}\Delta(B_s B_r^{-1};q,y) \ .
  \end{split}
\end{equation}
Since
\begin{equation}
  \begin{split}
    \prod_{s\in\{i_\alpha\}}\prod_{j=1}^{2N}\Delta(B_{i_\alpha} A_j^{-1};q,y)
   & =\frac{\prod_{i,j=1}^{2N}\Delta(B_iA_j^{-1};q,y)}{\prod_{s\in\overline{\{i_\alpha\}}}\Delta(B_sA_j^{-1};q,y)} \\
&    =\prod_{i,j=1}^{2N}\Delta(B_iA_j^{-1};q,y)\prod_{s\in\overline{\{i_\alpha\}}}\Delta(y A_j B_s^{-1};q,y) \ ,
  \end{split}
\end{equation}
one can immediately verify the equality of the left-hand-side and right-hand-side of \eqref{app:eq:fundSD}, and hence we have proven \eqref{app:eq:fundSD}.

\bigskip
Now, the integral \eqref{eq:fundSD} can be written as
\begin{equation}
\begin{split}
   \CI(a,b,c,d;q,y)
  =&\bDelta_{2-r_a-r_b}(a_ib^{-1}_j)\bDelta_{2-r_c-r_d}(c_id^{-1}_j)\\
   &\times\CI_{\rm SD}(\{y^{-\frac{r_a}{2}}a_i,y^{-\frac{r_c}{2}}c_i\},
           \{y^{\frac{r_b}{2}}b_i,y^{\frac{r_d}{2}}d_i\};q,y ) \ .
\end{split}
\end{equation}
The requirement of pole position \eqref{app:eq:polestructure} is satisfied if
\begin{equation}
  \begin{split}
    &|a_i|=|b_i|=|c_i|=|d_i|=1 \ ,\quad\quad i=1,\cdots,N \ ,\\
    &|q|<1,\quad |y|<1,\quad 0\leq|r|\leq1 \ .
  \end{split}
\end{equation}
Using \eqref{app:eq:fundSD} we see
\begin{equation}
  \begin{split}
     &\CI_{\rm SD}(\{y^{-\frac{r_a}{2}}a_i,y^{-\frac{r_c}{2}}c_i\},
           \{y^{\frac{r_b}{2}}b_i,y^{\frac{r_d}{2}}d_i\};q,y)\\
    &=\prod_{i,j=1}^{N}\Delta(y^{\frac{r_a+r_b}{2}}b_j a_i^{-1};q,y)\Delta(y^{\frac{r_a+r_d}{2}}d_ja_i^{-1};q,y)
    \Delta(y^{\frac{r_c+r_b}{2}}b_jc_i^{-1};q,y)\Delta(y^{\frac{r_c+r_d}{2}}d_jc_i^{-1};q,y)\\
     &\qquad\times\CI_{\rm SD}(\{y^{\frac{r_a-1}{2}}a^{-1}_i,y^{\frac{r_c-1}{2}}c^{-1}_i\},
           \{y^{\frac{1-r_b}{2}}b^{-1}_i,y^{\frac{1-r_d}{2}}d^{-1}_i\};q,y)\\
    &=\bDelta_{r_a+r_b}(b_ja^{-1}_i)\bDelta_{r_a+r_d}(d_ja^{-1}_i)\bDelta_{r_c+r_b}(b_jc^{-1}_i)\bDelta_{r_c+r_d}(d_jc^{-1}_i)\\
    &\qquad\times \CI_{\rm SD}(\{y^{\frac{r_a-1}{2}}a^{-1}_i,y^{\frac{r_c-1}{2}}c^{-1}_i\},
           \{y^{\frac{1-r_b}{2}}b^{-1}_i,y^{\frac{1-r_d}{2}}d^{-1}_i\};q,y) \ .
  \end{split}
\end{equation}
Using the identities
\begin{equation}
  \bDelta_{2-r_a-r_b}(a_ib^{-1}_j)\bDelta_{r_a+r_b}(b_ja^{-1}_i)=1 \ ,
\end{equation}
we arrive at \eqref{eq:fundSD}.

\subsection{Consistency of Residue Prescription with Gluing}

Since the Yang-Baxter duality is a sequence of the 2d Seiberg-like duality (applied four times),
the fundamental identity proven above should automatically imply the YBE.
The only subtlety here is that (as already commented in section \ref{R_from_2d}) 
the contour prescription obscures \eqref{eq.gauging}.
While \eqref{eq.gauging} is expected to hold on physical grounds, it would still be 
desirable to check explicitly that
the resulting integral is independent of the order of integration
of the two sides of the YBE.
For simplicity of the presentation we specialize to the case $N=1$,
where the contour prescription simplifies.

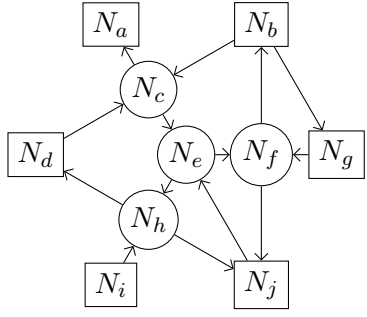
\begin{figure}[htbp]
  \centering
  \begin{tikzpicture}
    \node[F] (1) at (1,0){$N_i$};
	\node[F] (2) at (3,0) {$N_j$};
    \node[G] (3) at (1.5,0.866) {$N_h$};
    \node[F] (4) at (0,1.732) {$N_d$};
    \node[G] (5) at (2,1.732) {$N_e$};
    \node[G] (6) at (3,1.732) {$N_f$};
    \node[F] (7) at (4,1.732) {$N_g$};
    \node[G] (8) at (1.5,2.598) {$N_c$};
    \node[F] (9) at (1,3.464) {$N_a$};
    \node[F] (10) at (3,3.464) {$N_b$};
	\draw[->] (1)--(3);
    \draw[->] (3)--(2);
    \draw[->] (3)--(4);
    \draw[->] (5)--(3);
    \draw[->] (2)--(5);
    \draw[->] (6)--(2);
    \draw[->] (5)--(6);
    \draw[->] (7)--(6);
    \draw[->] (4)--(8);
    \draw[->] (8)--(5);
    \draw[->] (6)--(10);
    \draw[->] (10)--(7);
    \draw[->] (8)--(9);
    \draw[->] (10)--(8);
  \end{tikzpicture}
  \caption{\label{fig:YBquiver}The quiver for the YBE, compare with Figure \ref{fig.R_glue}.}
\end{figure}

Let us consider the quiver in Figure \ref{fig:YBquiver}. This is a quiver with three fundamental quivers (Figure \ref{fig.2dSeiberg_quiver}) glued together. Part of its index can be written is
\begin{equation}
\label{app:eq:consistencycheck}
\begin{split}
  \int\cdots\int[dc]\int[de]\, &\bDelta_{r_{ca}}\left(\frac{c}{a}\right)
  \bDelta_{r_{ce}}\left(\frac{c}{e}\right)
  \bDelta_{r_{bc}}\left(\frac{b}{c}\right)
  \bDelta_{r_{dc}}\left(\frac{d}{c}\right)
  \bDelta_{r_{ef}}\left(\frac{e}{f}\right)
  \bDelta_{r_{eh}}\left(\frac{e}{h}\right)
  \bDelta_{r_{je}}\left(\frac{j}{e}\right)\cdots,
\end{split}
\end{equation}
where $[dc]$ and $[de]$ are shorthand notation for combination of measure and vector multiplet contribution, and $\bDelta_{r_{ca}}\left(\frac{c}{a}\right)$ is the shorthand notation for
\begin{equation}
  \bDelta_{r_{ca}}\left(\frac{c}{a}\right)
  :=\prod_{\alpha,\beta=1}^N\left(y^{\frac{r_{ca}}{2}}\frac{c_\alpha}{a_\beta};q,y\right) \ .
\end{equation}
We keep only the part related to node $c$ and node $e$ in the index.

We would like to set
\begin{equation}
  q<y^\frac{r}{2}<1 \ .
\end{equation}
Hence $y^\frac{r}{2}a_\alpha$ and $y^\frac{r}{2}a_\alpha q^n$ with positive integer $n$ are always in the unit circle and $y^\frac{r}{2}a_\alpha q^{-n}$ are always outside the unit circle. On the other hand, $y^{-\frac{r}{2}}a^{-1}_\alpha$ and $y^{-\frac{r}{2}}a^{-1}_\alpha q^{-n}$ are outside the unit circle and $y^{-\frac{r}{2}}a^{-1}_\alpha q^n$ are inside the unit circle. Remember we always put the flavor fugacity $a_\alpha$ on the unit circle.

Now let us look at the integral \eqref{app:eq:consistencycheck}. For simplicity we look at $U(1)$ case. If we integrate over $c$ first, the only contribution are from the poles at
\begin{equation}
  y^{\frac{r_{bc}}{2}}b\  ,\quad y^{\frac{r_{dc}}{2}}d  \ ,
\end{equation}
and the integral becomes
\begin{equation}
  \begin{split}
    &\int\cdots\int[de]\, \bDelta_{r_{ef}}\left(\frac{e}{f}\right)
  \bDelta_{r_{eh}}\left(\frac{e}{h}\right)
  \bDelta_{r_{je}}\left(\frac{j}{e}\right)\\
    \times&\left[\bDelta_{r_{bc}+r_{ca}}\left(t^{\frac{r_{bc}+r_{ca}}{2}}\frac{b}{a}\right)
    \bDelta_{r_{bc}+r_{ce}}\left(\frac{b}{e}\right)
    \bDelta_{r_{dc}-r_{bc}}\left(\frac{d}{b}\right)
    +\bDelta_{r_{dc}+r_{ca}}\left(\frac{d}{a}\right)
    \bDelta_{r_{dc}+r_{ce}}\left(\frac{d}{e}\right)
    \bDelta_{r_{bc}-r_{dc}}\left(\frac{b}{d}\right)
    \right],
  \end{split}
\end{equation}
We then integrate over $e$ and pick up the residue at
\begin{equation}
  y^{\frac{r_{je}}{2}}j \ ,\quad y^{\frac{r_{bc}+r_{ce}}{2}}b
  \  ,\quad y^{\frac{r_{dc}+r_{ce}}{2}}d \ ,
\end{equation}
we get
\begin{equation}
  \begin{split}
    \int\cdots&
    \left[
    \bDelta_{r_{bc}+r_{ca}}\left(\frac{b}{a}\right)
    \bDelta_{r_{dc}-r_{bc}}\left(\frac{d}{b}\right)
    \bDelta_{r_{bc}+r_{ce}+r_{ef}}\left(\frac{b}{f}\right)
    \bDelta_{r_{bc}+r_{ce}+r_{eh}}\left(\frac{b}{h}\right)
    \bDelta_{r_{je}-r_{ce}-r_{bc}}\left(\frac{j}{b}\right)
    \right.\\
    &+\left.
    \bDelta_{r_{bc}+r_{ca}}\left(\frac{b}{a}\right)
    \bDelta_{r_{dc}-r_{bc}}\left(\frac{d}{b}\right)
    \bDelta_{r_{bc}+r_{ce}-r_{je}}\left(\frac{b}{j}\right)
    \bDelta_{r_{je}+r_{ef}}\left(\frac{j}{f}\right)
    \bDelta_{r_{je}+r_{eh}}\left(\frac{j}{h}\right)
    \right.\\
    &+\left.\bDelta_{r_{dc}+r_{ca}}\left(\frac{d}{a}\right)
    \bDelta_{r_{bc}-r_{dc}}\left(t^{\frac{r_{bc}-r_{dc}}{2}}\frac{b}{d}\right)
    \bDelta_{r_{dc}+r_{ce}+r_{ef}}\left(\frac{d}{f}\right)
    \bDelta_{r_{dc}+r_{ce}+r_{eh}}\left(\frac{d}{h}\right)
    \bDelta_{r_{je}-r_{dc}-r_{ce}}\left(\frac{j}{d}\right)
    \right.\\
    &+\left.
    \bDelta_{r_{dc}+r_{ca}}\left(\frac{d}{a}\right)
    \bDelta_{r_{bc}-r_{dc}}\left(\frac{b}{d}\right)
    \bDelta_{r_{dc}+r_{ce}-r_{je}}\left(\frac{d}{j}\right)
    \bDelta_{r_{je}+r_{ef}}\left(\frac{j}{f}\right)
    \bDelta_{r_{je}+r_{eh}}\left(\frac{j}{h}\right)
    \right].
  \end{split}
\end{equation}

Now let us integrate over $e$ first, we have poles at
\begin{equation}
  y^{\frac{r_{ce}}{2}}c\  ,\quad y^{\frac{r_{je}}{2}}j \ ,
\end{equation}
and the result is
\begin{equation}
  \begin{split}
        &\int\cdots\int[dc]\bDelta_{r_{ca}}\left(\frac{c}{a}\right)
  \bDelta_{r_{bc}}\left(\frac{b}{c}\right)
  \bDelta_{r_{dc}}\left(\frac{d}{c}\right)\\
    \times&\left[\bDelta_{r_{ce}+r_{ef}}\left(\frac{c}{f}\right)
    \bDelta_{r_{ce}+r_{eh}}\left(\frac{c}{h}\right)
    \bDelta_{r_{je}-r_{ce}}\left(\frac{j}{c}\right)
    +\bDelta_{r_{je}+r_{ef}}\left(\frac{j}{f}\right)
    \bDelta_{r_{je}+r_{eh}}\left(\frac{j}{h}\right)
    \bDelta_{r_{ce}-r_{je}}\left(\frac{c}{j}\right)
    \right],
  \end{split}
\end{equation}
then we integrate over $c$. It might seem that we have an extra pole at
\begin{equation}
  y^{\frac{r_{je}-r_{ce}}{2}}j \ ,
\end{equation}
however, the residue is zero because $\bDelta_{r_{je}-r_{ce}}\left(\frac{j}{c}\right)$ and $\bDelta_{r_{ce}-r_{je}}\left(\frac{c}{j}\right)$ terms have exactly opposite contribution and cancel out, hence we consider only the poles at
\begin{equation}
  y^{\frac{r_{bc}}{2}}b \ ,\quad
  y^{\frac{r_{dc}}{2}}d \ .
\end{equation}
The result is
\begin{equation}
  \begin{split}
    \int\cdots&
    \left[
    \bDelta_{r_{bc}+r_{ca}}\left(\frac{b}{a}\right)
    \bDelta_{r_{dc}-r_{bc}}\left(\frac{d}{b}\right)
    \bDelta_{r_{bc}+r_{ce}+r_{ef}}\left(\frac{b}{f}\right)
    \bDelta_{r_{bc}+r_{ce}+r_{eh}}\left(\frac{b}{h}\right)
    \bDelta_{r_{je}-r_{ce}-r_{bc}}\left(\frac{j}{b}\right)
    \right.\\
    &+\left.\bDelta_{r_{dc}+r_{ca}}\left(\frac{d}{a}\right)
    \bDelta_{r_{bc}-r_{dc}}\left(\frac{b}{d}\right)
    \bDelta_{r_{dc}+r_{ce}+r_{ef}}\left(\frac{d}{f}\right)
    \bDelta_{r_{dc}+r_{ce}+r_{eh}}\left(\frac{d}{h}\right)
    \bDelta_{r_{je}-r_{ce}-r_{dc}}\left(\frac{j}{d}\right)
    \right.\\
    &+\left.
    \bDelta_{r_{bc}+r_{ca}}\left(\frac{b}{a}\right)
    \bDelta_{r_{dc}-r_{bc}}\left(\frac{d}{b}\right)
    \bDelta_{r_{je}+r_{ef}}\left(\frac{j}{f}\right)
    \bDelta_{r_{je}+r_{eh}}\left(\frac{j}{h}\right)
    \bDelta_{r_{ce}-r_{je}+r_{bc}}\left(\frac{b}{j}\right)
    \right.\\
    &+\left.
    \bDelta_{r_{dc}+r_{ca}}\left(\frac{d}{a}\right)
    \bDelta_{r_{bc}-r_{dc}}\left(\frac{b}{d}\right)
    \bDelta_{r_{je}+r_{ef}}\left(\frac{j}{f}\right)
    \bDelta_{r_{je}+r_{eh}}\left(\frac{j}{h}\right)
    \bDelta_{r_{dc}+r_{ce}-r{je}}\left(\frac{d}{j}\right)
    \right].
  \end{split}
\end{equation}
The result is exactly the same as the previous result. We have proved explicitly the order of integration does not affect the result under residue prescription in $U(1)$ case.


\section{Dimensional Reduction of $\Delta(a;q,t)$} \label{sec:2d1d}

Let us discuss the dimensional reduction of the 2d index to the 1d index.
The discussion here is similar to the reduction of the $S^3 \times S^1$ partition function to
$S^3$ \cite{Dolan:2011rp,Gadde:2011ia,Imamura:2011uw}.

When the radius of the thermal cycle shrinks to zero, we expect
all the fugacities to approach to an identity. They can nevertheless
approach to an identity with a different scaling limit, first let us rewrite the fugacities as
\begin{equation}
  q=e^{2\pi i\tau} \ ,\quad y=e^{2\pi i\zeta}\ ,\quad a=e^{2\pi i\upsilon_a}.
\end{equation}
Dimension reduction means we set $\tau=i\beta$ then take the limit $\beta\rightarrow 0$, and also scale $\zeta$ and $\upsilon_a$ with  $\beta$,
\begin{equation}
  \zeta=\beta z \ ,\quad\upsilon_a=\beta\bm{a} \ ,
\end{equation}
hence
\begin{equation}
  \begin{split}
    \Delta(a;q,y)
    =&\prod_{n=0}^{\infty}\frac{(1-ya q^{n+1})(1-y^{-1}a^{-1}q^n)}{(1-a q^{n+1})(1-a^{-1}q^n)}\\
    =&\prod_{n=0}^{\infty}\frac{(1-e^{-2\pi \beta(-i(\bm{a}-z)+n+1)})(1-e^{-2\pi\beta(i(\bm{a}-z)+n)})}
                               {(1-e^{-2\pi\beta(-i\bm{a}+n+1)})(1-e^{-2\pi\beta(i\bm{a}+n)})} \ .
  \end{split}
\end{equation}
In the limit $\beta\rightarrow0$,
\begin{equation}
  \begin{split}
     \lim_{\beta\rightarrow0}\Delta(a;q,y)
    =&\frac{\pi(\bm{a}-z)}{\pi\bm{a}}
      \prod_{n=1}^{\infty}\frac{(\-i(\bm{a}-z)+n)(-(i\bm{a}+z)+n)}{(-i\bm{a}+n)(i\bm{a}+n)}\\
    =&\frac{\pi(\bm{a}-z)}{\pi \bm{a}}
      \prod_{n=1}^{\infty}\frac{1+\frac{(\bm{a}-z)^2}{n^2}}{1+\frac{\bm{a}^2}{n^2}}
    =\frac{\sinh\pi(\bm{a}-z)}{\sinh\pi \bm{a}} \ .
  \end{split}
\end{equation}

Similarly one can show that under the limit $\beta\rightarrow 0$,
\begin{equation}
  \lim_{\beta\rightarrow 0}\frac{\eta(q)^3}{i\theta_1(y^{-1};q)}
  =-\frac{1}{2i\beta}\frac{1}{\sinh\pi z} \ .
\end{equation}
This means that
under the dimension reduction,
the 2d index of a theory with $U(N)$ gauge group and fundamental chirals becomes,
\begin{equation}
  \left(-\frac{1}{2i\beta}\frac{1}{\sinh\pi z}\right)^N \int d\bm{a}\prod_{i\neq j}\frac{\sinh\pi(\bm{a}_i-\bm{a}_j)}{\sinh\pi((\bm{a}_i-\bm{a}_j)-z)}
  \prod_{i}\frac{\sinh\pi(\bm{a}_i+\left(\frac{r}{2}-1\right)z)}{\sinh\pi(\bm{a}_i+\frac{r}{2}z)} \ .
\end{equation}
This result, up to a divergent factor proportional to $1/\beta$, matches with the 1d index obtained in \cite{Hori:2014tda,Cordova:2014oxa}.

\bibliographystyle{jhep}
\bibliography{2dindex}

\end{document}